\documentclass[letterpaper]{no_journal}

\usepackage{graphicx}
\usepackage{comment}
\usepackage{xcolor}
\usepackage{booktabs}
\usepackage{xspace}

\usepackage{longtable}
\usepackage{makecell}

\journalyear{}
\journalvolume{}
\journalissue{}
\journalpages{}
\articledoi{10.15346/hc.v9i1.106}
\journalcopyright{\copyright{} 2022, Posch \& Bleier \& Fl{\"o}ck \& Lechner \& Kinder-Kurlanda \& Helic \& Strohmaier. CC-BY-3.0}

\title{Characterizing the Global Crowd Workforce: \\ A Cross-Country Comparison of Crowdworker Demographics}

\author{Lisa Posch\affil{Graz University of Technology}
        \and Arnim Bleier\affil{GESIS - Leibniz Institute for the Social Sciences}
        \and Fabian Fl{\"o}ck\affil{GESIS - Leibniz Institute for the Social Sciences}
        \and Clemens M. Lechner\affil{GESIS - Leibniz Institute for the Social Sciences}
        \and Katharina Kinder-Kurlanda\affil{University of Klagenfurt}        
        \and Denis Helic\affil{Graz University of Technology}
        \and Markus Strohmaier\affil{University of Mannheim, GESIS - Leibniz Institute for the Social Sciences \& Complexity Science Hub Vienna}}

\authorrunning{L. Posch et al.}

\newcommand{\amt}{MTurk\xspace}

\DeclareRobustCommand{\firstsecond}[2]{#1}

\usepackage{url}

\usepackage{breakurl}
\usepackage{placeins}

\begin{document}

\maketitle

\begin{abstract}
Since its emergence roughly a decade ago, microtask crowdsourcing has been attracting a heterogeneous set of workers from all over the globe. This paper sets out to explore the characteristics of the international crowd workforce and offers a cross-national comparison of crowdworker populations from ten countries. We provide an analysis and comparison of demographic characteristics and shed light on the significance of microtask income for workers situated in different national contexts. 
With over 11,000 individual responses, this study is the first large-scale country-level analysis of the characteristics of workers on the platform Appen (formerly CrowdFlower and Figure Eight), one of the two platforms dominating the microtask market. 
We find large differences between the characteristics of the crowd workforces of different countries, both regarding demography and regarding the importance of microtask income for workers. Furthermore, we find that the composition of the workforce in the ten countries was largely stable across samples taken at different points in time.
\end{abstract}

\section{Introduction}

Doing freelance work over an online platform or by other digital means is a mode of labor that has existed since at least the 1990s.
However, a specific type of online freelance work, called microtask crowdsourcing, has only gained increasing popularity with an international workforce over roughly the past 15 years and has no offline equivalent. Microtask crowdsourcing is characterized by an anonymous crowd doing freelance work for many different, largely unknown principals, whereby the freelance work constitutes working on very small-scale, non-expert tasks that are continuously available from central platforms.\footnote{Other modes of crowdsourcing exist, such as winner-takes-all contests for ideas or more macro-size tasks, essentially constituting online freelancing (cf. \citealp{kuek2015global, pongratz2018crowds, bhattacharyya2018studying}).} Microtask crowdsourcing, through its low-barrier nature, offers potential income opportunities for almost everyone with an Internet connection.

Precise estimates of the number of platforms, their users, and turnovers for this type of work are hard to come by, as no official labor market statistics for crowdwork exist as of yet, and proprietary platforms seldom release such information. However, experts postulate a significant and lasting growth of microtask platforms, assuming a market size of \$500 million in 2016, with the amount of global microtask workers being put at around 9 million, up from 4 million in 2013 \cite{kuek2015global}. 
The World Bank \cite{kuek2015global}, the European Agency for Health and Safety at Work \cite{euroagency2015} and other official bodies have in recent years been discussing chances and perils of this new form of income for millions of people, and they see the need for better regulation, but also plainly for better insights into the crowdsourcing market. 
Scholars and legislators have, for instance, expressed qualms about the tendency of crowdwork -- often meant to offer supplementary income -- to evolve into a main income source for workers in precarious economic circumstances, while at the same time being unregulated, volatile in terms of pay and availability, not offering union-typical bargaining powers, and requiring predominantly monotonous work (see, e.g., \citealp{kuek2015global, waas2017crowdwork}). On the flip side, observers have highlighted the opportunities that crowdwork offers, especially for inhabitants of regions with sub-par working conditions in ``offline'' employment \cite{kuek2015global}.

To inform this discussion around the impact of crowdwork on communities around the world, research concerned with the demographic composition of the international crowd workforce is very valuable, 
not least to enable comparisons with the more traditional, offline workforce and with the general population (see, e.g., \citealp{paolacci2014inside, weinberg2014comparing, huff2015these}). 
In this regard, it is also strongly linked with the study of \textit{why} crowdworkers are attracted to this new form of employment  (e.g., \citealp{chen2019more, mcms_published, brewerwould, brabham2010threadless}). 
Furthermore, the demographic characteristics of the crowd workforce may influence regulatory responses regarding crowdwork \cite{waas2017crowdwork}.

Demographic information is also instrumental for optimizing the use of crowd platforms as recruiting instruments to infer knowledge about a broader ground population or at least control for sampling biases -- e.g., for using crowdworkers as an affordable and expeditious alternative for sampling participants in social and behavioral science research \cite{paolacci2014inside}. Lastly, it is valuable for understanding task performance linked to demographic features, e.g., for labeling, translation, or speech recognition tasks (e.g., \citealp{kazai2012face, pavlick2014language}).

While its useful applications seem apparent, knowledge about the demographic composition of the crowd workforce remains spotty. 
Out of the two major microtask platforms dominating the market,\footnote{\amt and CrowdFlower have been estimated to share 80\% of all revenue generated in the microtask market, with revenues approximately equal \cite{kuek2015global}.} only the demographic composition of the predominantly American and Indian crowdworkers on Amazon Mechanical Turk (\amt) is sufficiently well-known
(e.g., \citealp{ipeirotis2010demographics, ross2010crowdworkers, berg2016income, difallah2018demographics}), but insights about other platforms -- and particularly workers in countries outside the \amt target audiences -- are few and far between.

To gain insights into the global crowd workforce, it is important to analyze microtask platforms other than \amt, as the vast majority of \amt's workforce consists of workers located in the United States and India (see, e.g., \citealp{difallah2018demographics}). The country distribution on \amt is likely due to the restrictive payment options that the platform offers in other countries.\footnote{Before 2019, workers could receive money from
\amt in the USA and India while workers from other countries were paid in Amazon.com gift
cards \cite{amtpay}. In 2019, \amt enabled payments in US\$ for workers in 25 countries outside the U.S., provided that they owned a U.S. bank account \cite{amazon_payment_change}. However, this change seems to have done little to attract a more international workforce to \amt, as current data from mturk tracker \cite{difallah2018demographics} indicates. Data from mturk tracker is available at \url{http://demographics.mturk-tracker.com}.} By contrast, parts of the payment process of the second microtask market leader, CrowdFlower (since 2019 known as Appen and formerly also known as Figure Eight)\footnote{The platform's name changed from CrowdFlower to Figure Eight in 2018, and in 2019, Figure Eight was acquired by the company Appen \cite{appen_press}. At the time of our data collection, which started in 2016, the platform's name was CrowdFlower. For consistency with the survey questions, we therefore refer to the platform as CrowdFlower, rather than Appen or Figure Eight, in the remainder of this paper.}, are handled by independent partner websites,
which provides workers with significantly more flexibility regarding their payment. This is likely the reason that CrowdFlower attracts a much more international workforce than \amt.

This paper therefore sets out to complement the existing literature by mapping out the demographics of workers on CrowdFlower, exploring its international workforce to shed light on cross-national differences. 
We conducted a survey of CrowdFlower workers in ten countries, over two time points, collecting information about their demographic characteristics and about the centrality of microtasks in their lives, regarding time spent as well as importance and use of microtask income.

The main contributions of this paper are (1) a large-scale comparison of crowdworker demographics in ten different countries, (2) a comparison of the centrality of microtasks in the workers' lives in these ten countries, and (3) an analysis of the changes in these features between two samples taken eight months apart. The results of this study advance our understanding of the international crowd workforce and provide important insights for researchers and policy makers seeking to understand this new form of work.

The paper is structured in the following way. Section~\ref{sec:rw} gives an overview of related work on the characteristics of crowdworkers. Section~\ref{sec:survey} describes our survey design and the process of data collection. In Section~\ref{sec:demographics}, we present a cross-national comparison of crowdworker demographics, and Section~\ref{sec:importance} presents a comparison of the importance that microtasks have for workers in different countries. Section~\ref{sec:discussion} provides a discussion of the results of this study, regarding the differences between countries and regarding the stability of our findings. Finally, Section~\ref{sec:conclusion} concludes this paper.

\section{Related Work}
\label{sec:rw}

Most of the research investigating demographic and economic characteristics of workers on microtask platforms has focused on the platform Amazon Mechanical Turk (\amt). Early studies on the demographics of workers on \amt \cite{ipeirotis2010demographics,ross2009turkers, ross2010crowdworkers,paolacci2010running, kazai2012face} found that the vast majority of workers were located in the USA and India, and that they were young and highly educated. Workers were predominantly female in the USA and predominantly male in India. A small but significant percentage of workers relied on \amt to make basic ends meet.

Later studies on the demographics of \amt workers reported similar results (e.g., \citealp{goodman2017crowdsourcing, berg2016income, peer2017beyond, pavlick2014language, naderi2018crowdworkers, difallah2018demographics}). 
On \amt, American and Indian crowdworkers still constitute the vast majority of workers, together accounting for over 80\% of the worker population on \amt.\footnote{Also see \url{http://demographics.mturk-tracker.com/\#/countries/all}.}
Consistent with earlier studies, \citet{berg2016income} found that Indian and American workers on \amt were young and well-educated. Indian workers were predominantly male, but there was now more gender balance among workers from the U.S. These findings are also supported by current data collected by \emph{mturk tracker}\footnote{http://www.mturk-tracker.com} \cite{ipeirotis2010analyzing, difallah2018demographics}. 
\citet{pavlick2014language} conducted a study on the languages spoken by bilingual workers on \amt and found that the majority of workers who accepted their translation tasks were located in either the USA or India. Nevertheless, there were sufficient bilingual workers to accurately and quickly complete translation tasks for 13 different languages.

Research on the demographics of workers on \amt is closely linked with questions concerning the representativeness of \amt samples and their suitability for different research purposes (e.g., \citealp{goodman2017crowdsourcing}).
For example, \citet{paolacci2010running} compared American crowdworkers on \amt to the general U.S. population and found that workers in the USA were more representative of the population than university subject pools. Compared to the general U.S. population, crowdworkers tended to be slightly younger and, despite being more highly educated, workers had a lower income level. This observation could be partially explained by age. 
\citet{buhrmester2011amazon} compared \amt workers to standard Internet samples. Their \amt sample was more diverse than both standard Internet samples and American college samples. They found that \amt workers were similar in gender distribution, more non-white, almost equally non-American, and older than the standard Internet sample. 
\citet{berinsky2012evaluating} evaluated the suitability of crowdworker samples for experimental political science and found that the respondents recruited on \amt were more representative of the U.S. population than in-person convenience samples, but less representative than respondents recruited for Internet-based panels or national probability samples. Furthermore, they found that crowdworkers responded to experimental stimuli in a way that was consistent with prior research. 

\citet{weinberg2014comparing} analyzed socio-demographic characteristics of workers on \amt and compared them to the characteristics of respondents of a population-based web panel. They found that the \amt participants were younger and more educated, and there was a higher proportion of women than among the web panel participants. The \amt sample was more divergent from the general population than the web panel.
\citet{huff2015these} analyzed the demographics and political characteristics of \amt workers from the United States and compared them to the respondents of the Cooperative Congressional Election Survey (CCES), a stratified sample survey conducted yearly in the United States. They found that \amt was, in many cases, good at attracting those demographics that were difficult to attract for CCES (e.g., young Asian males). Furthermore, they found that the distribution of employment in different occupational sectors of workers on \amt was very similar to that of CCES respondents and that the respondents were located in similar locations on the rural-urban continuum. 

\citet{shapiro2013using} investigated the suitability of crowdworker samples for conducting research on psychopathology, investigating the prevalence of different psychiatric disorders and related problems among crowdworkers on \amt. They concluded that \amt might be useful for studying clinical and subclinical populations.
\citet{paolacci2014inside} analyzed the characteristics of \amt as a participant pool for the social sciences and concluded that worker samples from \amt could replace or supplement convenience samples in psychological research, but that they should not be considered representative of a country's population.

Research on the demographics of workers on other microtask platforms, and therefore also on workers based in countries other than the USA and India, is more scarce. Furthermore, due to reasons such as unavailability of demographic data beyond the workers' location or small sample sizes, none of these studies have so far analyzed and compared the demographics of workers at the country level.

\citet{hirth2011human} analyzed the demographics of the platform Microworkers with respect to the home countries of requesters and workers and found that the platform was much more geographically diverse than \amt. The countries with the largest number of workers were Indonesia, Bangladesh, India, and the United States, accounting for 60\% of the workforce on the platform. The remaining 40\% were dispersed over a heterogeneous set of geographical locations. Using the United Nations Human Development Index \cite{hdi} to categorize countries, they found that an almost equal proportion of workers were located in countries with low development (24\%) and countries with very high development (21\%), while the majority of the workforce was located in countries with medium development (45\%).  

\citet{martin2017understanding} compared the demographics of workers on \amt to the demographics of workers on the platforms Microworkers and Crowdee. The study grouped workers on the Microworkers platform into workers located in ``Western countries'' (including all workers from Europe, Oceania, and North America) and workers located in ``developing countries'' (including all workers from South America, Asia, and Africa). The results of their demographic survey indicated that workers on Microworkers and Crowdee were predominantly male, younger than workers on \amt, and highly educated. A large proportion of workers reported working either full-time or part-time on all three platforms. Regarding the differences between ``Western countries'' and ``developing countries,'' the study found that workers in the ``developing countries'' group were younger, lived in larger households, were more educated, had a lower household income, and spent more time on the platform, compared to workers in the ``Western countries'' group.

\enlargethispage{1\baselineskip}

Further, few studies have concerned themselves with the workforce demographics of the platform CrowdFlower, which covers half the market share for microtasks and traditionally employs a geographically diverse set of workers. 

\citet{berg2016income} collected demographic data from a geographically diverse sample of workers on CrowdFlower and found that only 2.8\% of workers (10 respondents) were from the U.S. and 8.5\% were from India (30 respondents). The workers on CrowdFlower were predominantly male and more educated than American workers on \amt, but less educated than Indian workers on \amt.
\citet{mubarak2016demographic} conducted a survey of Arab workers on CrowdFlower at two points in time. 
At each time point, they collected 500 responses and found that Arab workers were predominantly male ($\geq 75\%$), that most workers were between 20 and 39 years old, and that they were highly educated. 
The two points in time were about 1.5 months apart and yielded very similar results, with the largest difference being that the workers were slightly older in the second data collection. 
\citet{peer2017beyond} examined the demographics of workers on  CrowdFlower and Prolific Academic and compared them to the demographics of workers on \amt.
The study found that, compared to \amt, both CrowdFlower and Prolific Academic had a higher proportion of male workers, and the mean age was similar on all three platforms. CrowdFlower had the highest diversity in terms of race, and both CrowdFlower and Prolific Academic were much more diverse in worker location than \amt. On all three platforms, workers were highly educated.

In sum, there have been extensive studies on the characteristics and demographic composition of American and Indian workers on \amt, whereas research on the characteristics of crowdworkers on other microtask platforms and on the demographic composition and characteristics of workers in countries other than the USA and India remains sparse. In comparison, we provide a survey of workers pre-selected to cover similar respondent numbers for ten diverse countries, over two time points, whereas previous studies have studied samples that were not stratified by locations and have mostly not controlled for temporal changes. In doing so, we have conducted the most comprehensive scientific collection of worker characteristics on CrowdFlower to date, with 11,946 individual responses collected (after spam removal). Using the data collected with our survey, we provide the first country-level comparison of (i) the demographics and (ii) the centrality of microtasks in the lives of crowdworkers on this platform and show that notable differences exist between countries.

\section{Survey}
\label{sec:survey}

In order to provide insights into the characteristics of the international crowd workforce, we conducted a large survey in ten different countries on the CrowdFlower platform. To gain insights into the stability of the distributions of the different characteristics, we repeated the survey after eight months.

\begin{table}[b!]
\caption{\textbf{Sample sizes and percentage of spam received.} This table shows the sample sizes of the different groups at both time points, as well as the percentage of spam received. $\mathbf{N}_{raw}$ shows the total number of responses collected, before removing spam. $\mathbf{N}_{T1}$ and $\mathbf{N}_{T2}$ show the number of responses after spam removal in sample \texttt{T1} and \texttt{T2}, respectively. $\mathbf{Spam}_{T1}$ and $\mathbf{Spam}_{T2}$ show the percentage of workers who did not pass all attention checks, for \texttt{T1} and \texttt{T2}.}
\centering
\begin{tabular}{l l c c c c c} 
\toprule
\textbf{Group} & \textbf{Code} & $\mathbf{N}_{raw}$ & $\mathbf{Spam}_{T1}$ & $\mathbf{N}_{T1}$ & $\mathbf{Spam}_{T2}$ & $\mathbf{N}_{T2}$  \\  
 \midrule
 All & ALL & 18000 & 35 \% & 5857 & 32\% & 6089 \\
 
 \midrule
 
 High Income & HIGH & 5400 & 28 \% & 1952 & 26\% & 1988 \\
 Middle Income & MID & 5400 & 32 \% & 1834 & 31\% & 1863 \\
 Low Income & LOW & 5400 & 44 \% & 1508 & 38\% & 1679 \\
 \midrule

 USA & USA & 1800 & 20 \% & 721 & 14\% & 776 \\
 Spain & ESP & 1800 & 25 \% & 677  & 30\% & 634 \\
 Germany & DEU & 1800 & 38 \% & 554 & 36\% & 578 \\%[1mm]
 
 Brazil & BRA & 1800 & 45 \% & 496 & 43\% & 509 \\
 Russia & RUS & 1800 & 25 \% & 677 & 21\% & 708 \\%[1mm]
 Mexico & MEX & 1800 & 27 \%  & 661 & 28\% & 646 \\

 India & IND & 1800 & 32 \% & 608 & 28\% & 645 \\
 Indonesia & IDN & 1800 & 55 \% & 401  & 47\% & 476 \\
 Philippines & PHL & 1800 & 45 \% & 499  & 38\% & 558 \\%[1mm]
 
 Venezuela & VEN & 1800 & 37 \% & 563 & 38\% & 559 \\
 \bottomrule
\end{tabular}
\label{table:demographic}
\end{table}

\subsection{Data Collection}
We posted the survey as a microtask on CrowdFlower.
The task included seven questions about workers' demographics and three questions about the centrality of microtasks in workers' lives. Furthermore, the task contained questions about workers' motivation for putting effort into microtasks, which were used for the validation of the Multidimensional Crowdworker Motivation Scale \cite{mcms_published}. Anonymity was assured in the task instructions.

We collected data of workers from ten countries, with 900 participants in each country at each time point. 
In our country selection, we aimed for diverse income levels by selecting countries from three different World Bank income groups.\footnote{The World Bank country classification for the time of our data collection is available at \url{https://databank.worldbank.org/data/download/site-content/OGHIST.xls}. Here, we use the group label ``Middle Income'' (MID) for the upper middle income group and ``Low Income'' (LOW) for the lower middle income group for better readability.} Furthermore, we aimed for a high cultural diversity and sufficient activity on CrowdFlower.\footnote{The country had to either be high in the Alexa (\url{http://www.alexa.com/}) ranking or one of the top contributing countries in at least one of CrowdFlower's partner channels.} 
The countries that we selected for the high income group were USA, Germany, and Spain, for the middle income group we selected Brazil, Russia, and Mexico, and for the low income group we selected India, Indonesia, and the Philippines. Additionally, we collected data from Venezuela because it was the most active country on CrowdFlower at the time of the start of the data collection.\footnote{CrowdFlower received 18.5\% of its traffic from Venezuela at the time of data collection. This data was obtained from \url{http://www.alexa.com/.}} However, Venezuela represents a special case concerning income due to the circumstance that the black market exchange rate deviates from the official exchange rate to a large extent \cite{blackmarket}. Therefore, we did not assign Venezuela to any of our income groups.

We posted the survey on CrowdFlower at different times during the day and the week in order to capture a diverse sample of workers and to account for fluctuations in worker composition by the hour of the day and the day of the week.\footnote{There are indications that worker composition varies by time of the day and day of the week (see, e.g., \url{http://demographics.mturk-tracker.com}).} For each country, 300 responses were requested during typical working hours (8:00 am to 5:00 pm in the appropriate time zone), 300 responses were requested in the evening (6:00 pm to 11:00 pm in the appropriate time zone), and 300 responses were requested during weekends.\footnote{Specifically, we posted the survey as a task on CrowdFlower, once for each country and each time point. We initially requested 300 responses for each task, and we then extended each task twice, requesting 300 additional responses for each extension during the respective time intervals.}
After the first round of data collection, which took place in October and November of 2016 (\texttt{T1}), we conducted a second round of data collection in June and July of 2017 (\texttt{T2}).\footnote{To ensure that our findings do not depend heavily on seasonal effects, we chose to conduct the second round of data collection after eight months (as opposed to a full year).}

The survey was open to workers of all CrowdFlower levels.
In both rounds of data collection, we used the default set of CrowdFlower channels, and workers were able to participate in the survey once in each round of data collection. Before being able to submit the task, workers had to have it open for a minimum of 150 seconds. To minimize population bias in the responses, we aimed for a payment similar to most other tasks on the platform. Based on previous research on average earnings on microtask platforms (e.g., \citealp{kaplan2018striving, berg2016income,ross2010crowdworkers,horton2010labor, khanna2010evaluating}) and on the first author's experience as a crowdworker on CrowdFlower, we paid US\$0.1 for the task. After completing a task on CrowdFlower, the platform asks workers to judge the task payment relative to other tasks (\emph{``How would you rate the pay for this task relative to other tasks you've completed?''}) on a five-point scale ranging from ``much worse'' to ``much better''. For our survey tasks, we received average ratings ranging from 3.3 (in Germany, \texttt{T2}) to 4.1 (in Mexico and India, for both time points), which indicates that workers perceived our task payment as equal to or ``somewhat better'' than other tasks.

The survey was conducted in English in all countries. While this approach only captures crowdworkers with sufficient English skills, demand for crowdworkers is driven by Anglophone clients, and English is the dominant language in task requests \cite{kuek2015global}. Congruently, English is expected by CrowdFlower to be spoken by all workers at a sufficient level to solve tasks, as made apparent by its interface language and English being assumed a guaranteed language skill for all workers in the platform's worker language selection settings. The survey presented by CrowdFlower to workers after completing a task included a question about the clarity of the task instructions and interface. For our task, workers rated the task instructions and interface as ``clear'' to ``very clear'' in most countries, and as ``somewhat clear'' to ``clear'' in Indonesia (both time points) and Brazil (\texttt{T2}). This indicates that most workers had little difficulty understanding the English task instructions and interface. Furthermore, the responses to the motivation scale that was included in the task provided strong evidence that the workers understood English well \cite{mcms_published}.

\begin{table}[b!]
\caption{\textbf{Survey Questions.} This table shows the survey questions in our CrowdFlower task. A detailed overview of the survey questions and answer options can be found in Appendix~\ref{app:answer_options}.}
\centering
\begin{tabular}{p{5mm} l c} 
  \\
\toprule
  \textbf{Demographics} & \\
  D1 & \it{What is your gender?}\\ 
  D2 & \it{What is your age?}\\ 
  D3 & \it{What is your marital status?}\\ 
  D4 & \it{How many people live in your household?}\\ 
  D5 & \it{What is your highest education level?}\\ 
  D6 & \it{What is your employment status (CrowdFlower tasks excluded)?}\\ 
  D7 & \it{What is your approximate household income, per year (after taxes, in US\$)?}\\ 
\midrule
  \textbf{\mbox{Importance of Microtasks}} & \\
  I1 & \it{How much time do you spend on CrowdFlower, per week?}\\ 
  I2 & \it{Is the money from CrowdFlower your primary source of income?} \\
  I3 & \it{What do you do with the money that you earn on CrowdFlower?} \\[.5mm] 

 \bottomrule
\end{tabular}
\label{table:survey_questions}
\end{table}

\begin{table}[b!]
\caption{\textbf{Categories of Money Use.} For each category, this table shows exemplary responses to the open-ended survey question ``What do you do with the money that you earn on CrowdFlower?''.}
\centering
\begin{tabular}{p{14.5cm}} 
  \\
\toprule
  \textbf{Basic Expenses} \\
  ``I buy food!!'' (USA), ``I use the money to help pay my monthly rent.'' (USA), ``buy sensors to glucose measure'' (Spain), ``I use it for my daily needs like to pay rent and buy my essentials.'' (India), ``use it for my medicines'' (India) \\
\midrule
  \textbf{Leisure Activities}  \\
  ``I will use to pay for my hobbies.'' (USA), ``entertainment, eating out'' (USA), ``Go to the cinema.'' (Spain), ``With the money I usually do trips.'' (Spain), ``Use it as pocket money'' (India) \\
\midrule
  \textbf{Save/Invest}  \\  
  ``Put it in a savings account'' (USA), ``Build a BitCoin investment portfolio'' (USA), ``I keep it for the future'' (Spain)'', ``I save all the money I earn'' (Spain), ``i save the money for investments'' (India), ``saving for marriage and future life'' (India) \\
\midrule
  \textbf{Buy Gifts}  \\  
  ``I will save it to try and afford a gift for my children for christmas'' (USA), ``spend it on xmas presents for my kid'' (USA), ``small Gifts'' (Spain), ``buy gifts to my four daughters'' (Spain), ``Use it to buy gifts for my children.'' (India), ``used it for my mom dad's anniversary'' (India) \\
\midrule
  \textbf{Education}  \\  
    ``Pay for my college tuition.'' (USA), ``Use it for my driving test.'' (Spain), ``Save it [t]o pay for my college expenses'' (Spain), ``The Money I have Earned in CrowdFlower is Used for my Studies.'' (India), ``for further studies'' (India) \\ 
\midrule
  \textbf{Donate to Charity}  \\   
    ``I will spend for my family and remaining to charity.'' (India), ``I want to do lot of the things, primary is to donate a share out of it [...].'' (India), ``Almost 80\% paid for poor children's fee.'' (India), ``helping to poor peoples'' (India) \\
\midrule
 \textbf{Other}  \\ 
 ``Nothing yet, this is my first task.'' (USA), ``Multiple things, nothing in particular.'' (Spain), ``it is very small amount to spend i have not earned  so much'' (India), ``I have not much enough to withdraw it.'' (India)
 \\
 \bottomrule
\end{tabular}
\label{table:freetext_money_use}
\end{table}

In the tasks, we included four attention checks for detecting spam, such as workers clicking randomly or accepting the task despite having insufficient English skills. The use of attention checks is an established method to detect spam in survey research (see, e.g., \citealp{krosnick1999survey, oppenheimer2009instructional, vannette2014comparison, abbey2017attention, gummer2021using}), and attention checks have also been employed in crowdsourcing research design (e.g., \citealp{paolacci2010running}). 
Three of our attention check questions asked workers to pick a specific ranking on a 7-point scale. The fourth attention check question asked workers ``Are you paying attention to the questions?'', offering ``No,'' ``Yes,'' and ``I don't know'' as possible answers in a drop-down list. We removed the responses of all workers who did not pass all four attention check questions, ensuring that less than 0.1\%  ($(\tfrac{1}{7})^3*\tfrac{1}{3}$) of spammers passed the test if they answered all four questions at random.

Table \ref{table:demographic} shows the number of respondents per income group and country for each time point, as well as the percentage of spam received and the number of respondents after spam removal. As it is the crowdworkers' choice whether to accept a task or not, our samples are necessarily self-selected, as is generally the case for surveys on microtask platforms. 
Overall, we received slightly less spam in \texttt{T2} than in \texttt{T1}. However, this was not the case in all countries: Workers from seven countries produced less non-valid responses in \texttt{T2} than in \texttt{T1}, while three countries produced more spam in \texttt{T2}. These differences suggest that while the amount of spam received in a certain task might not be constant over time, the amount does appear to vary by country. For example, workers in Indonesia produced the highest number of non-valid responses in both time points (followed by Brazil and the Philippines), while workers in the U.S. produced the lowest number of non-valid responses in both time points. One reason for the differences might be that understanding the survey questions requires a higher cognitive effort from workers with lower levels of English skills, potentially leading to a higher temptation to pick an answer at random.

\enlargethispage{-1\baselineskip}

In the survey, we asked crowdworkers about seven demographic characteristics and about three aspects concerning the importance of microtasks in their life. Table \ref{table:survey_questions} shows the questions.
The question about crowdworkers' use of the money earned through microtasks (I3) was constructed as a multiple choice question. We aimed for a high-level distinction of money use to keep the number of answer options low (and reduce the total survey length). As standard survey instruments for capturing expenditures of households or individuals are very detailed in their classifications and do not provide canonical distinctions at a sufficiently high level for our purposes\footnote{Cf., e.g., the ``Consumer Expenditure Survey Interview Questionnaire'' \cite{CEX_surveys} or the "Classification of Individual Consumption according to Purpose" \cite{UN_coicop}. If coarser distinctions are defined, they are generally at the binary consumption vs. savings/other expenditures level (cf. \citealp{destatis2013einkommens}).} -- and to account for potential particularities of crowdworker money use patterns -- we opted for an inductive approach to constructing the answer options.

To this end, we posted a preceding open-ended survey task, where we asked workers the question \emph{``What do you do with the money that you earn on CrowdFlower?''} For answering the question, we provided workers with a free text field for their answer, which could be arbitrarily long. 
We posed this question to workers in the USA, Spain, and India.\footnote{For the development of the answer options, we used the same countries as for the development of the Multidimensional Crowdworker Motivation Scale \cite{mcms_published}. USA and India were selected because these countries have significant populations of crowdworkers on different platforms. Spain was selected in order to include a European country with a sufficiently large population of crowdworkers.} In each country, 300 workers were surveyed in October 2016. Two authors of this paper then manually categorized the open-ended responses. Workers often reported more than one use for the money earned through microtasks, so each answer could be coded with multiple categories.

We then used these manually identified categories to construct the answer options for the survey question \texttt{I3}: (1) \emph{I use the money for basic living expenses (food, rent, sanitary items, medical care,...)}, (2) \emph{I spend the money on leisure activities (hobbies, games, holidays, sports,...)}, (3) \emph{I save/invest the money}, (4) \emph{I use the money to buy gifts for other people}, (5) \emph{I use the money to finance my education}, (6) \emph{I donate the money to charity}, and (7) \emph{Other purposes}. Table~\ref{table:freetext_money_use} shows example answers for each category, along with the country the answers stem from.

\section{Demographics}
\label{sec:demographics}

In this section, we report the results of the demographics section of the survey (see questions D1-D7 in Table~\ref{table:survey_questions}). For each demographic characteristic, we report the proportion of each answer choice in the ten countries as an average of \texttt{T1} and \texttt{T2}, the differences in proportion between the countries and, per country, the differences in proportions between the two samples taken eight months apart. As a measure of difference, we report the Jensen-Shannon (JS) divergence \cite{lin1991divergence} between the respective answer distributions for each demographic characteristic. The JS divergence quantifies how dissimilar two distributions are and is bounded by $1$ and $0$. A value of $0$ indicates equivalence between the distributions, and higher values indicate the degree of dissimilarity. The reported JS divergences between two countries are the averages of the divergences between these countries at \texttt{T1} and \texttt{T2}. The results presented in this section are also available as an interactive visualization on our website \cite{crowdworkersinfo}.

\subsection{Gender}

\begin{figure}
  \centering
    \includegraphics[width=0.5\textwidth]{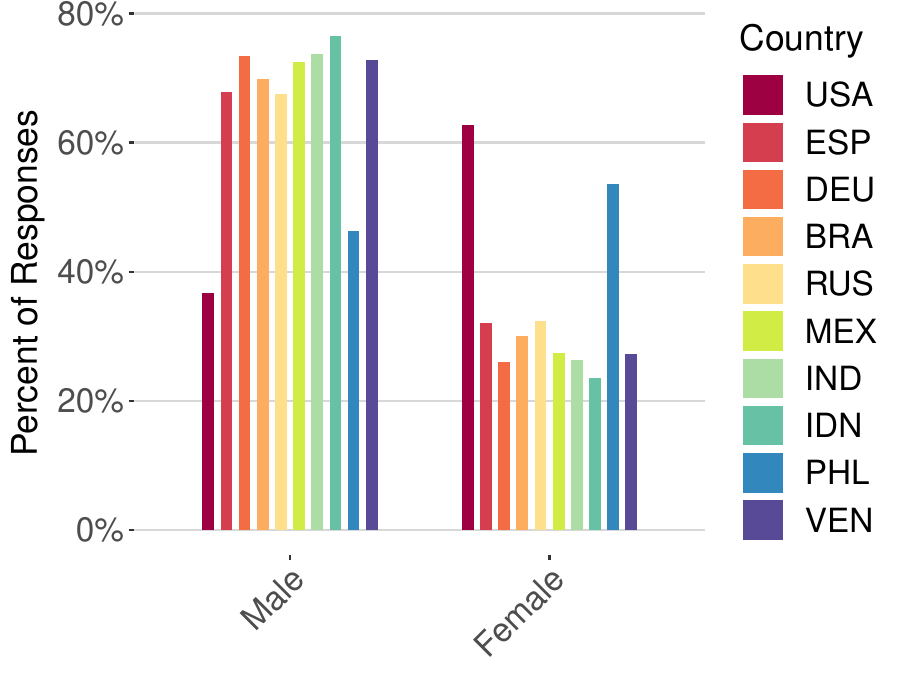} 
  \caption{\textbf{Gender Distribution.} This figure shows the gender distribution of workers in the different countries. The bar height represents the average of \texttt{T1} and \texttt{T2}. In most countries, crowdworkers were predominantly male. Only in the USA and the Philippines, female workers constituted the majority.}
  \label{fig:gender}
\end{figure}

In most countries, crowdworkers were predominantly male, with the proportion of male workers exceeding 60\%. The gender distribution was similar in all countries, with the exceptions of the USA and the Philippines, which were the only two countries where female workers constituted the majority. The most gender-balanced workforce was present in the Philippines, with 52\% (in \texttt{T1}) and 55\% (in \texttt{T2}) percent of workers being women. Figure~\ref{fig:gender} shows the gender distribution in the ten countries. The height of the bars corresponds to the average of the proportions at \texttt{T1} and \texttt{T2}. 
The answer options to the gender question included a third category, ``other,'' which is not included in Figure~\ref{fig:gender} due to the small number of responses. The differences between the sums of the male and female percentages and 100\% are due to this third category. 

\begin{figure}[t!]
  \centering
\includegraphics[width=0.8\textwidth]{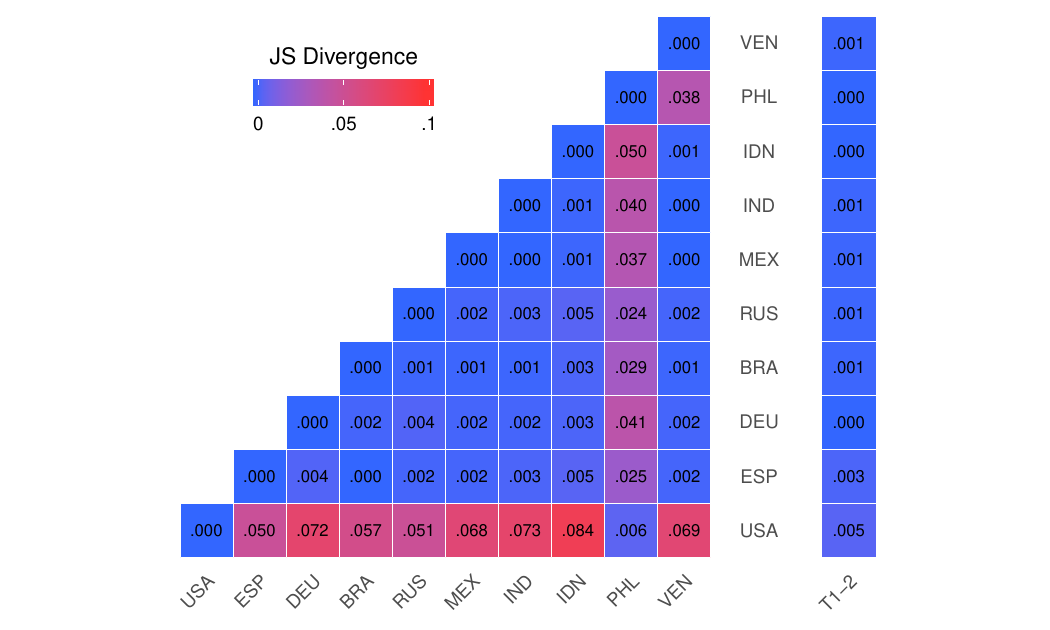}
  \caption{\textbf{Gender JS}. This figure shows the JS divergences between the gender distributions of the different countries. The bar on the right shows the JS divergence between \texttt{T1} and \texttt{T2} for each individual country.}
  \label{fig:gender_js}
\end{figure}

The gender distributions of American and Indian workers are consistent with findings of early studies on \amt (e.g., \citealp{ipeirotis2010demographics, paolacci2010running}) which found that in the United States, there were more female than male workers, and in India, there were more male workers. However, the United States crowd workforce on \amt, at the time of data collection, was more gender-balanced than the US-based crowd workforce on CrowdFlower.\footnote{For data on the gender distribution of American and Indian workers on \amt, see \url{http://demographics.mturk-tracker.com/\#/gender/all}.} 
\citet{ipeirotis2010demographics} hypothesized that this gender distribution difference between India and the United States may be due to the fact that in the United States, \amt is often used by stay-at-home parents and underemployed or unemployed workers (which are more likely to be female), while in India, workers are more likely to rely on \amt as a primary source of income. However, our results show that this does not generally hold true for the differences between the gender distribution of high income and low income countries. 

Figure~\ref{fig:gender_js} shows the JS divergences of the answer distributions between each country pair and, for each country, the divergence between \texttt{T1} and \texttt{T2}.
The gender distribution was mostly stable between the time points, with Spain and the USA exhibiting the largest differences in distributions. The divergence in the USA was mainly due to the gender category ``other,'' as which ten crowdworkers identified in \texttt{T2}, compared to none in \texttt{T1}. The divergence in Spain was due to an increased proportion of female workers in \texttt{T2}, where 35\% of workers reported being female in \texttt{T2} compared to 30\% in \texttt{T1}. While the change was less pronounced in other countries, the percentage of female crowdworkers slightly increased from \texttt{T1} to \texttt{T2} in all countries except Russia.

\subsection{Age}

Crowdworkers were young in all countries, with most crowdworkers being between 18 and 34 years of age. This is consistent with studies on \amt (e.g., \citealp{ipeirotis2010demographics, ross2010crowdworkers, berg2016income, difallah2018demographics}), which found that younger workers were overrepresented on the platform. 

\begin{figure}[t!]
  \centering
\includegraphics[width=0.9\textwidth]{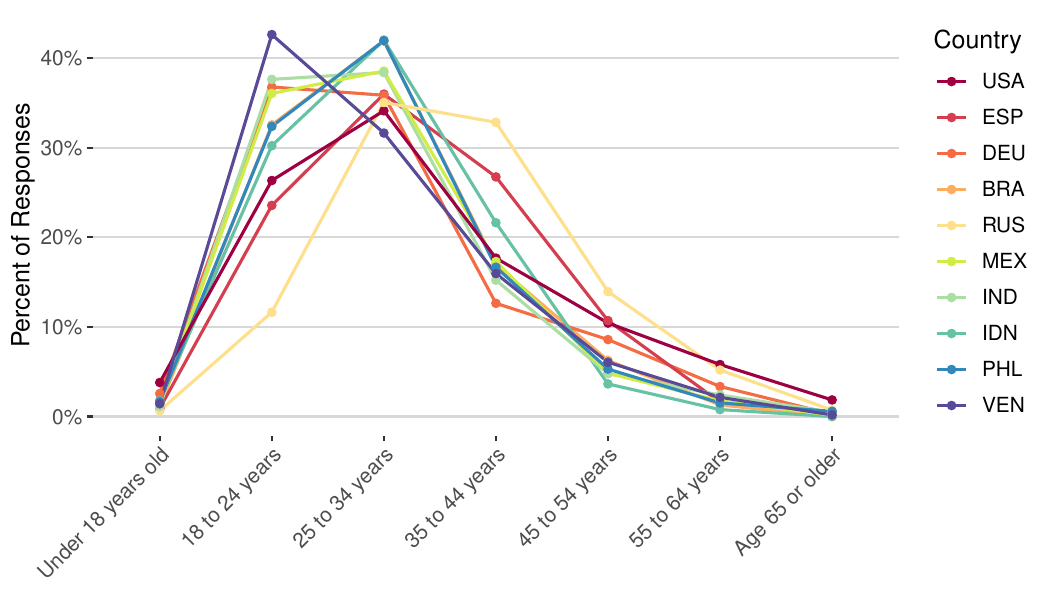}
\vspace*{-2mm}
  \caption{\textbf{Age Distribution}. This figure shows the age distribution of workers in the different countries. The percentages represent the average of \texttt{T1} and \texttt{T2}. Crowdworkers were generally young in all countries, with most crowdworkers being between 18 and 34 years of age. Russia was the country with the oldest population of crowdworkers, and Venezuela had the highest proportion of very young workers (aged 18-24).}
  \label{fig:age}
\end{figure}

The country with the oldest population of crowdworkers on CrowdFlower was Russia, which had by far the lowest proportion of workers aged between 18-24 years and the highest population of workers aged between 35 and 54 years old. Venezuela had the highest proportion of very young workers (aged 18-24). Figure~\ref{fig:age} shows the age distribution in the ten countries.
Data from mturk tracker\footnote{\url{http://demographics.mturk-tracker.com/\#/yearOfBirth/all}} indicates that Indian workers on \amt tend to be younger than American workers (also see \citealp{difallah2018demographics}). This difference also seems to be present on CrowdFlower, especially for the proportion of workers aged 18 to 24 years, which formed a much higher percentage of the Indian crowd workforce than the American crowd workforce.

\begin{figure}[h!]
  \centering
  \vspace*{6mm}
\includegraphics[width=0.8\textwidth]{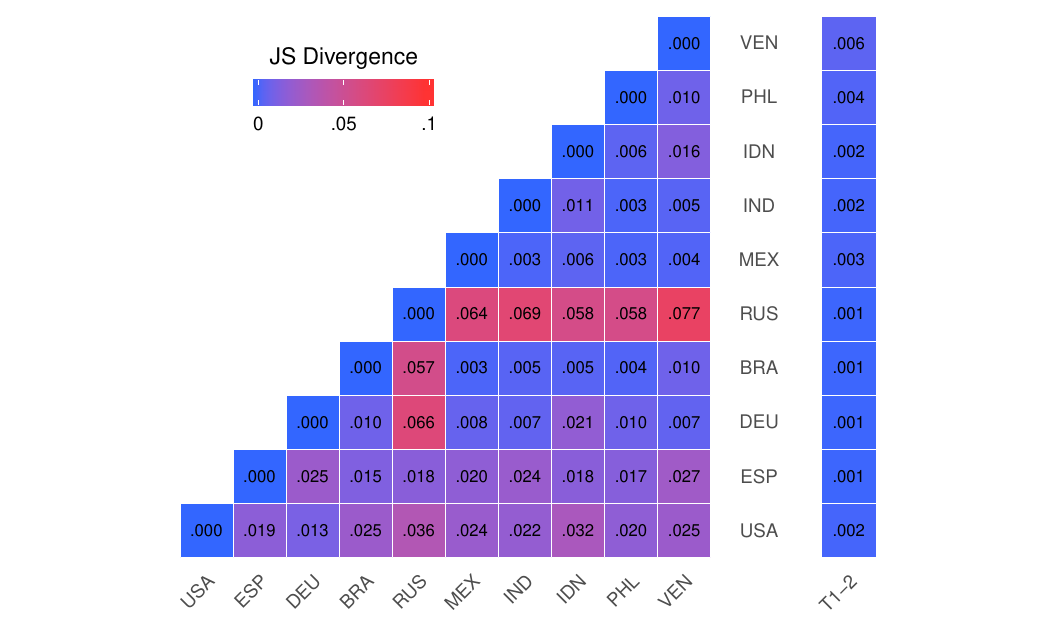}
\vspace*{-3mm}
  \caption{\textbf{Age JS}. This figure shows the JS divergences between the age distributions of the different countries. The bar on the right shows the JS divergence between \texttt{T1} and \texttt{T2} for each individual country.}
  \label{fig:age_js}
\end{figure}

Figure~\ref{fig:age_js} shows the JS divergences of the age distributions.
In most countries, there was little difference in age distribution between \texttt{T1} and \texttt{T2}. Venezuela had the largest difference in age distribution between the two time points, mostly due to an increase in young workers in the age bracket 18 to 24 years and a decrease in workers aged over 24.

\subsection{Marital Status}

\begin{figure}
  \centering
    \includegraphics[width=0.6\textwidth]{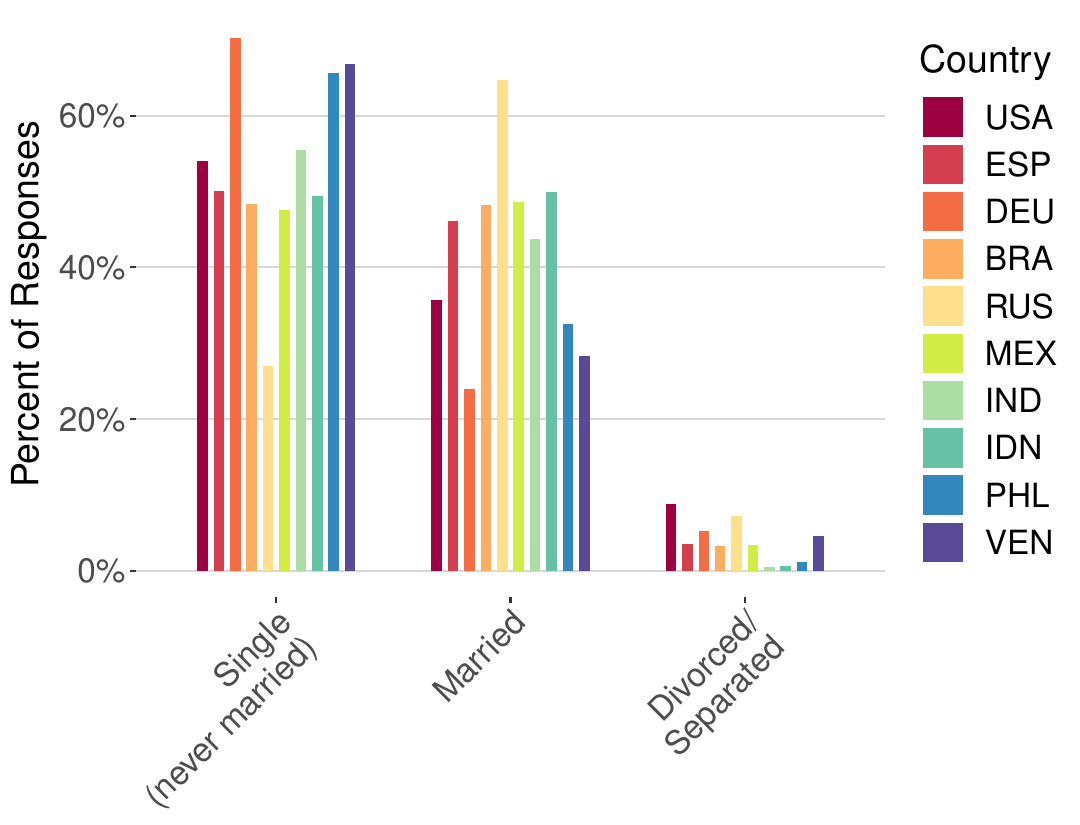}
  \caption{\textbf{Distribution of Marital Status.} This figure shows the marital status distribution of workers in the different countries. The bar height represents the average of \texttt{T1} and \texttt{T2}. Six countries had a higher proportion of non-married workers than married workers. Russia had the highest proportion of married crowdworkers, and it was the only country with more than 60\% married workers.}
  \label{fig:marital}
\end{figure}

\begin{figure}
  \centering
\includegraphics[width=0.8\textwidth]{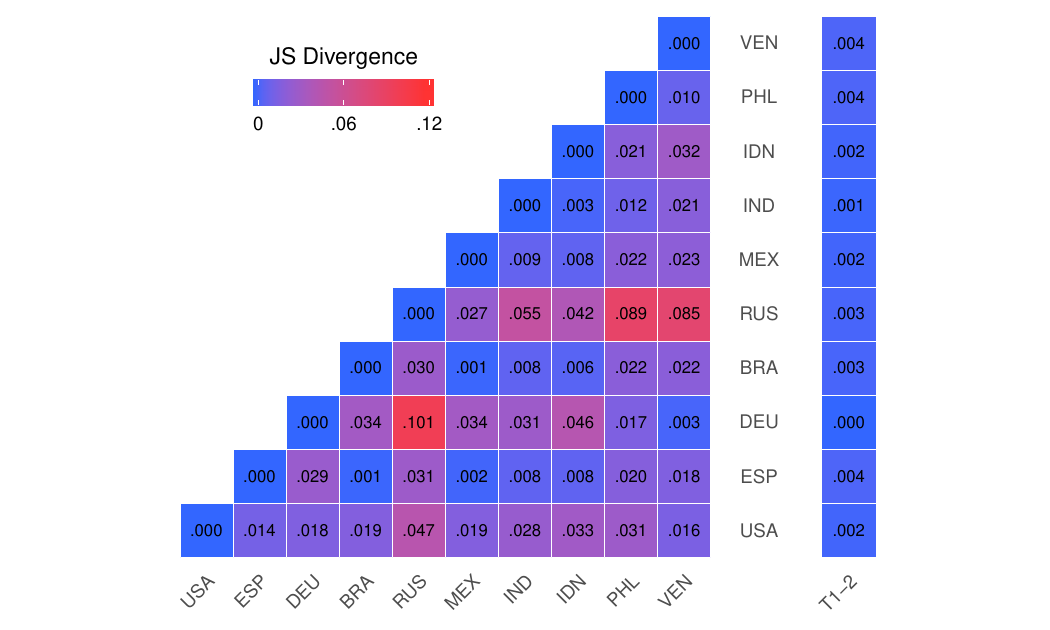}
  \caption{\textbf{Marital Status JS}. This figure shows the JS divergences between the marital status distributions of the different countries. The bar on the right shows the JS divergence between \texttt{T1} and \texttt{T2} for each individual country.}
  \label{fig:marital_js}
\end{figure}

Six countries had a higher proportion of non-married workers than married workers. Of all countries, Russia had the highest proportion of married crowdworkers, and it was the only country with more than 60\% married workers. USA and Russia were the countries with the largest proportion of divorced or separated workers. The countries with the highest proportion of non-married workers were Germany, the Philippines, and Venezuela. Figure~\ref{fig:marital} shows the distribution of the workers' marital status in the ten countries.
The response option for this survey question also included the category ``widowed,'' which received a very small number of responses and is therefore not included in Figure~\ref{fig:marital}.

The findings regarding American and Indian workers are somewhat consistent with findings from \amt.
On both CrowdFlower and \amt, Indian workers reported a higher proportion of married workers than U.S.-based workers (see, e.g., \citealp{difallah2018demographics}). However, on \amt, the difference was more pronounced than on CrowdFlower, and, in contrast to Indian workers on CrowdFlower, the majority of Indian workers on \amt who were surveyed by mturk tracker\footnote{\url{http://demographics.mturk-tracker.com/\#/maritalStatus/all}} reported being married.

Figure~\ref{fig:marital_js} shows the JS divergences of the answer distributions. Russia had the highest divergences with other countries due to the high proportion of married workers. Regarding the differences between the time points, in most countries, there was a slight decrease in the proportion of married workers from \texttt{T1} to \texttt{T2}. The only country where this was not the case was Germany, which also had the most stable distribution of marital status between the time points.

\subsection{Household Size}
 
\begin{figure}
  \centering
    \includegraphics[width=0.9\textwidth]{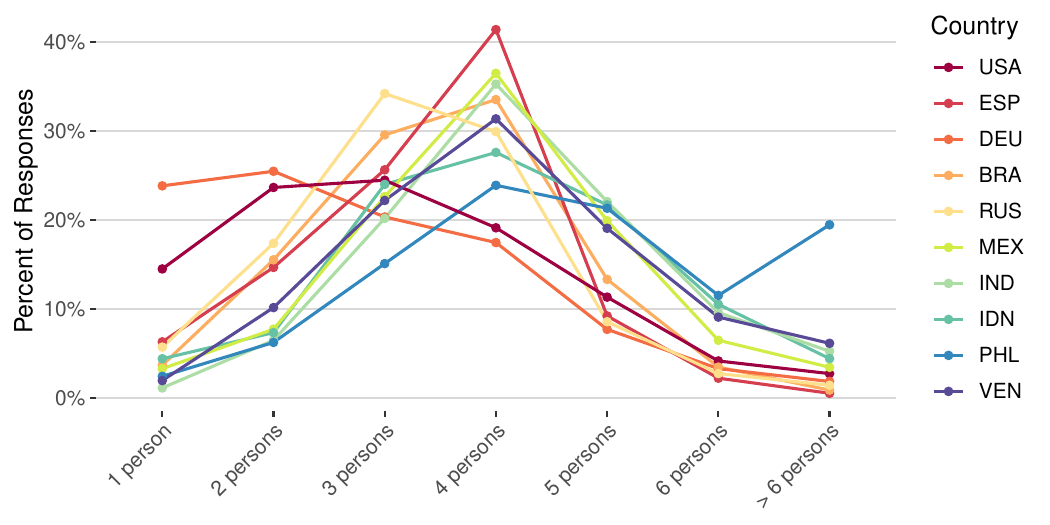}
  \caption{\textbf{Distribution of Household Size.} This figure shows the household size of workers in the different countries. The percentages represent the average of \texttt{T1} and \texttt{T2}. Germany had the highest proportion of single and two-person households, followed by the USA. All other countries had a very low proportion of single households (below 10\%). The Philippines was the country with by far the highest proportion of households with more than seven persons.}
  \label{fig:household_size}
\end{figure}

Germany had the highest proportion of single and two-person households, followed by the USA. All other countries had a very low proportion of single households (below 10\%). 
The Philippines was the country with by far the highest proportion of households with more than seven persons, with more than double the proportion of all other countries.
Spain had the highest proportion of four-people households, and Russia had the highest proportion of three-person households.
India's crowd workforce reported the lowest proportion of single households. 
Figure~\ref{fig:household_size} shows the distribution of household size.

Data from mturk tracker\footnote{\url{http://demographics.mturk-tracker.com/\#/householdSize/all}} shows that on \amt, workers in India tend to live in larger households than American workers. This difference in household size was also present in the workers on CrowdFlower. Workers located in the United States mainly lived in households with two or three persons, while a four-person household was the most common response among workers in India.

\begin{figure}
  \centering
\includegraphics[width=0.8\textwidth]{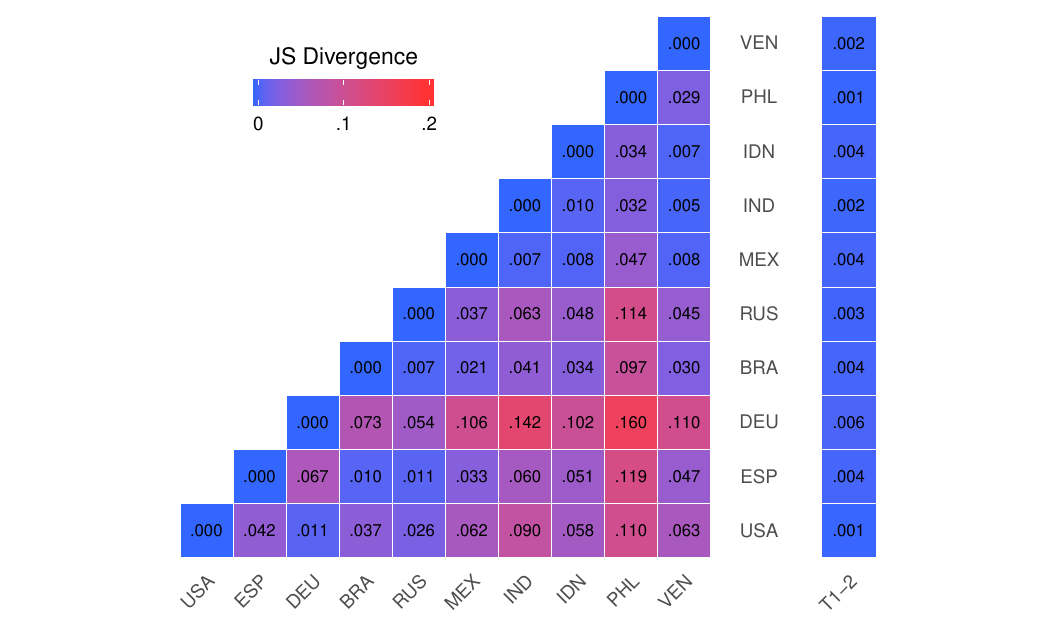}
  \caption{\textbf{Household Size JS}. This figure shows the JS divergences between the household size distributions of the different countries. The bar on the right shows the JS divergence between \texttt{T1} and \texttt{T2} for each individual country.}
  \label{fig:household_size_js}
\end{figure}

Figure~\ref{fig:household_size_js} shows the JS divergences of the household size distributions. Generally, we found the largest divergences between the countries of the high income group as well as Russia and the low income countries.
There were no large differences in household size distribution between the two time points. The German sample had the largest difference, with more workers reporting living in two-person households in \texttt{T2} than in \texttt{T1}, and fewer workers reporting three-person households.

\subsection{Employment Status}

\begin{figure}
  \centering
    \includegraphics[width=0.85\textwidth]{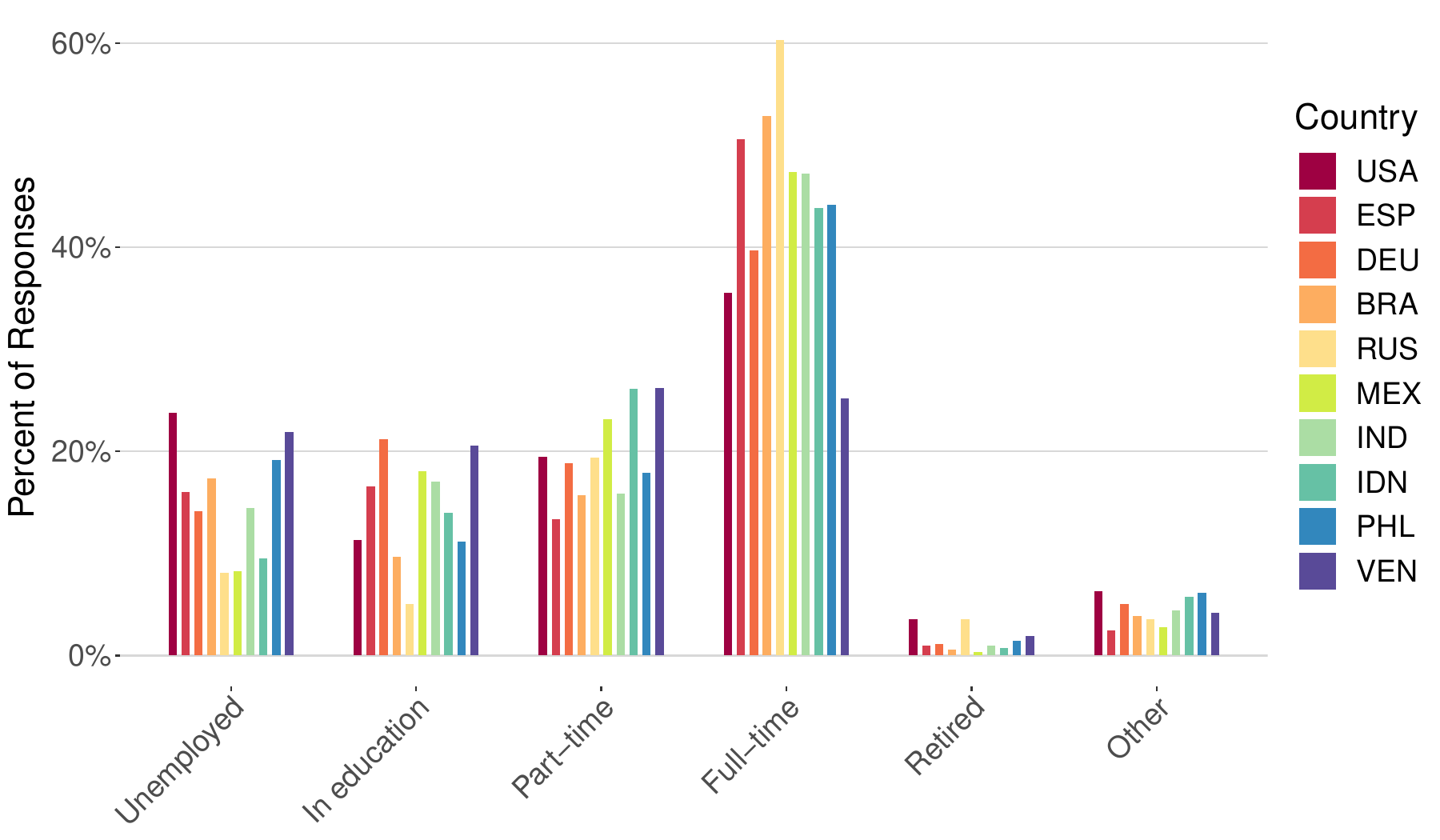}
  \caption{\textbf{Employment Status Distribution.} This figure shows the employment status distribution of workers in the different countries, CrowdFlower tasks excluded. The bar height represents the average of \texttt{T1} and \texttt{T2}. In almost all countries, over 35\% of workers had a full-time job besides their activity on CrowdFlower. The only exception to this was Venezuela.}
  \label{fig:employment_status}
\end{figure}

\begin{figure}
  \centering
\includegraphics[width=0.8\textwidth]{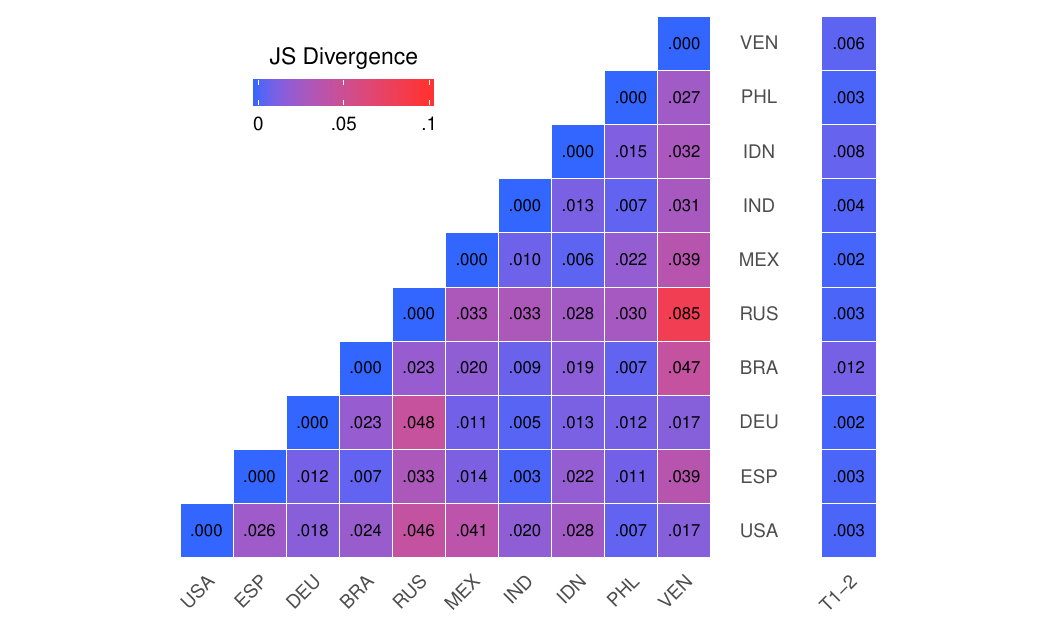}
  \caption{\textbf{Employment Status JS}. This figure shows the JS divergences between the employment status distributions of the different countries. The bar on the right shows the JS divergence between \texttt{T1} and \texttt{T2} for each individual country.}
  \label{fig:employment_status_js}
\end{figure}

The question regarding workers' employment status asked crowdworkers to explicitly exclude their activity on CrowdFlower. Figure~\ref{fig:employment_status} shows the distribution of employment status.
In almost all countries, over 35\% of workers had a full-time job besides their activity on CrowdFlower. The only exception to this was Venezuela, where only 28\% had full-time jobs at \texttt{T1} besides CrowdFlower. This percentage was even lower in \texttt{T2}, where only 23\% of Venezuelan workers reported having full-time jobs. 
A significant proportion of workers reported being in education, with Germany and Venezuela having the highest proportion of workers in education. 
The highest proportion of unemployed workers was reported in the United States, followed by Venezuela.
Very few workers reported being retired, which is very likely due to the overall young age of the workers. 

Similar numbers have been reported for workers on \amt. An early study on \amt \cite{ipeirotis2010demographics} found that around 30\% of American and Indian workers on \amt had either no job or only a part-time job besides crowdwork. However, \citet{berg2016income} later found that only 60\% of \amt workers held any job besides crowdwork, which is a number close to our findings on American and Indian workers on CrowdFlower.

Figure~\ref{fig:employment_status_js} shows the JS divergences between the employment status distributions. The largest difference in employment status distribution was between Russia and Venezuela. While Russian workers reported the highest percentage of workers in full-time employment, Venezuela reported the lowest percentage of all countries. Furthermore, there were large differences between Russia and Venezuela in the proportion of workers who reported being unemployed or in education.

In most countries, fewer workers reported working full-time in \texttt{T2} than in \texttt{T1}, while the percentage of unemployed workers, workers in education, and part-time workers increased from \texttt{T1} to \texttt{T2}. In Brazil, which had the largest JS divergence between the time points, this change was most pronounced, with a large decrease of workers in full-time employment (from 59.5\% in \texttt{T1} to 46.2\% in \texttt{T2}) and a large increase of unemployed workers (from 13.1\% in \texttt{T1} to 21.6\% in \texttt{T2}). An exception to this pattern was Germany, where the percentage of workers employed full-time stayed roughly the same, while there was a slight decrease in unemployed workers and a slight increase of workers in education. The second-largest JS divergence between time points was in Indonesia, where a lower proportion of workers reported having a full-time job in \texttt{T2} than in \texttt{T1}, and a higher proportion of workers reported holding a part-time job.

\subsection{Education Level}

\begin{figure}
  \centering
    \includegraphics[width=0.99\textwidth]{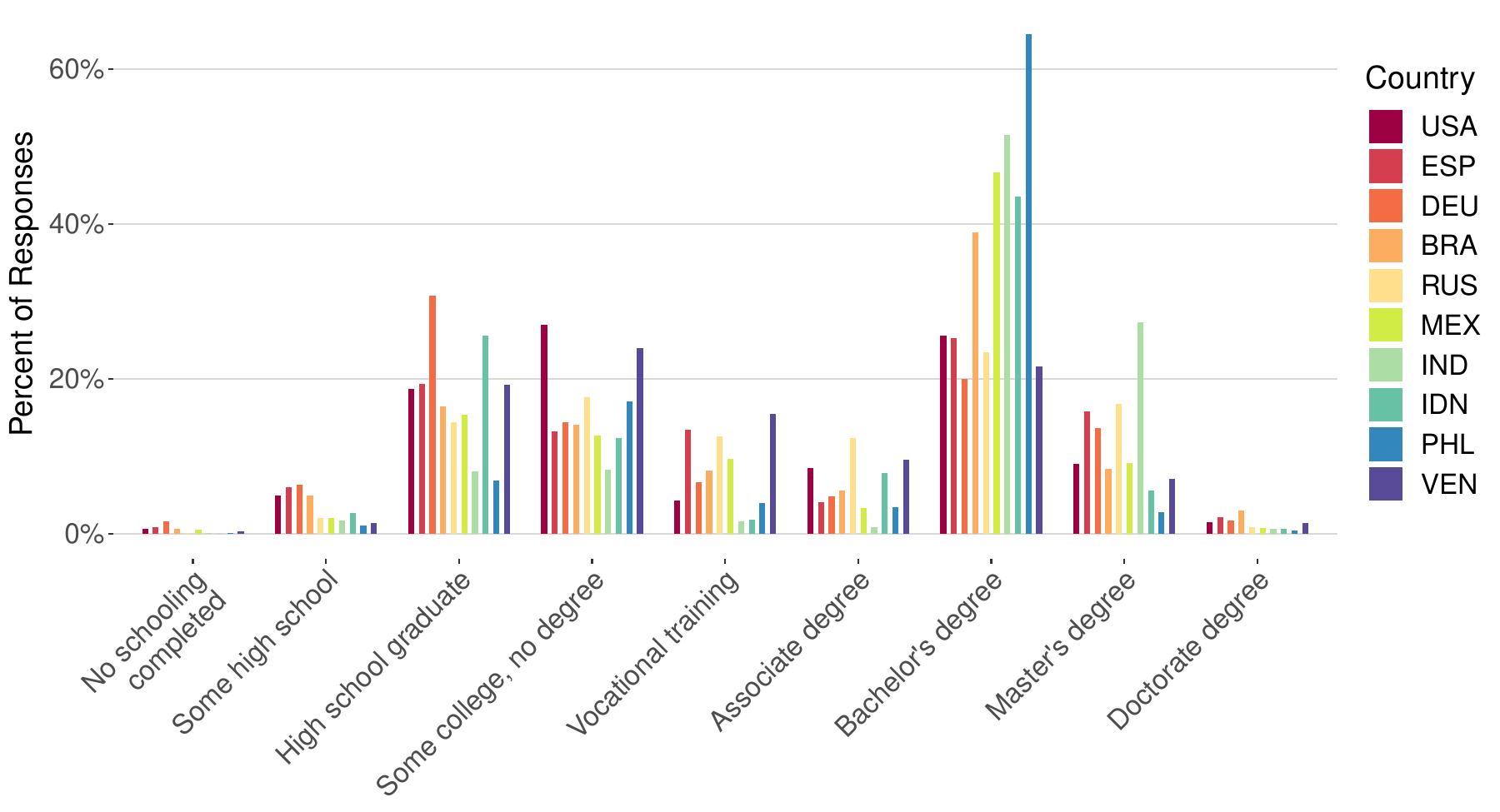}
  \caption{\textbf{Education Level Distribution.} This figure shows the education level distribution of workers in the different countries. The bar height represents the average of \texttt{T1} and \texttt{T2}. Crowdworkers are generally well educated, and the proportion of workers with at least a Bachelor's degree was 30\% or above in all countries. The countries with the highest proportion of college graduates were India and the Philippines.}
  \label{fig:education_level}
\end{figure}

Crowdworkers on CrowdFlower are generally well educated. The proportion of workers having a Bachelor's degree or higher was 30\% or above in all countries. Figure~\ref{fig:education_level} shows the distribution of education level.

Workers in the low income group countries reported especially high education levels. The countries with the highest proportion of college graduates were India and the Philippines. India also had the highest proportion of workers with a Master's degree of all countries. 

Our finding that crowdworkers are generally highly educated is consistent with the findings of studies on the demographics of \amt (e.g., \citealp{berg2016income}), and it contrasts with the notion that micro-work is especially attractive to unemployed people with no specialized skills (e.g., \citealp{kuek2015global}). The fact that workers from lower income countries tend to have higher education levels is consistent with the findings of studies on \amt (e.g., \citealp{ipeirotis2010demographics}), which found that Indian workers on \amt tend to have more education than workers from the United States. An exception to this pattern seems to be Venezuela, where workers tend to be less educated than in other low income countries.\footnote{While we did not include Venezuela in the low income country group due to the reasons stated in Section~\ref{sec:survey}, Venezuelan workers reported a very low household income.}

\enlargethispage{-1\baselineskip}

\begin{figure}
  \centering
\includegraphics[width=0.8\textwidth]{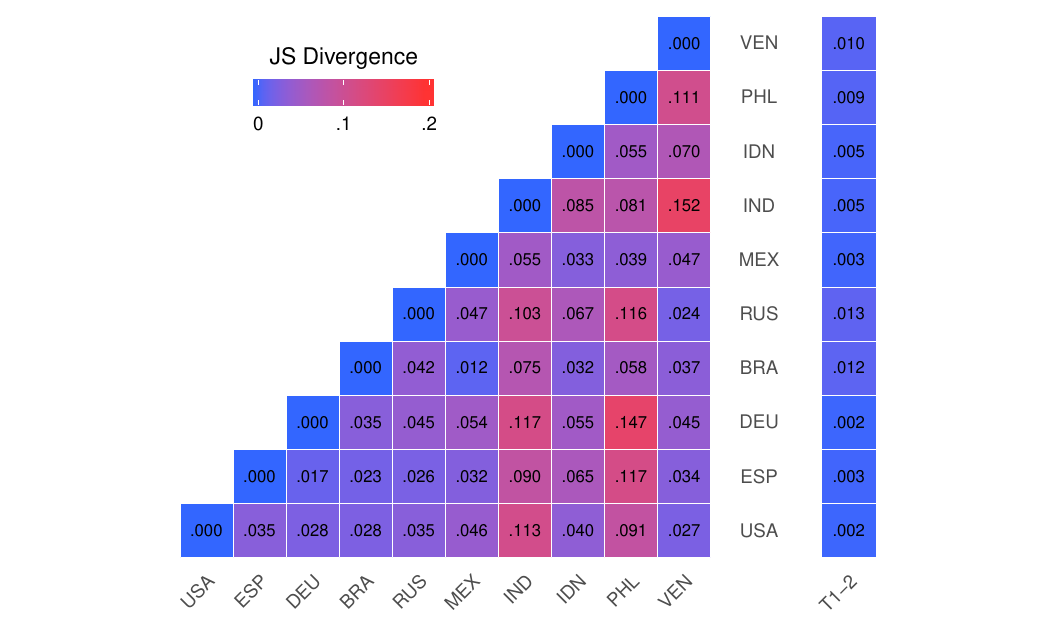}
  \caption{\textbf{Education Level JS}. This figure shows the JS divergences between the education level distributions of the different countries. The bar on the right shows the JS divergence between \texttt{T1} and \texttt{T2} for each individual country.}
  \label{fig:education_level_js}
\end{figure}

\FloatBarrier

Very few workers reported having no schooling completed at all (below 2\% in all countries), and only a small proportion of workers reported having only ``some high school'' education. 
Germany had the highest proportion of workers with a high school degree but no college education.

Figure~\ref{fig:education_level_js} shows the JS divergences between the education level distributions.
The largest difference in distribution was between Venezuela and India, with the proportion of Indian workers with a Bachelor's degree being more than twice as high and the proportion of workers with a Master's degree being over three times higher than the proportion in Venezuela.

\begin{figure}
  \centering
    \includegraphics[width=0.9\textwidth]{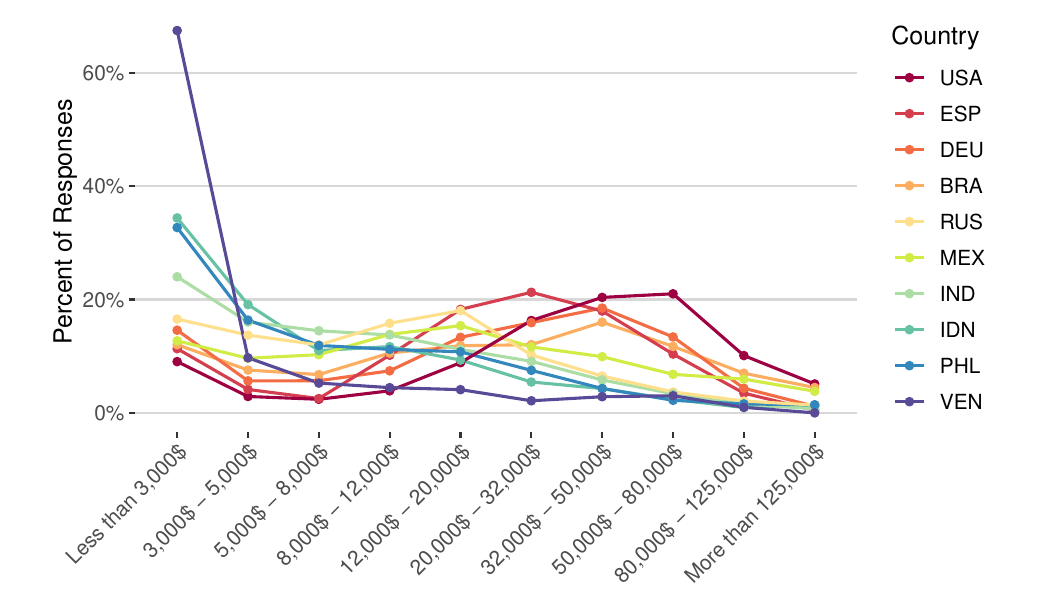}
  \caption{\textbf{Distribution of Household Income.} This figure shows the household income distribution of workers in the different countries. The percentages represent the average of \texttt{T1} and \texttt{T2}. Apart from Venezuela, the reported income distributions are largely consistent with the World Bank classification of the countries. Workers from Venezuela reported by far the lowest annual household income.}
  \label{fig:household_income}
\end{figure}

\begin{figure}
  \centering
\includegraphics[width=0.8\textwidth]{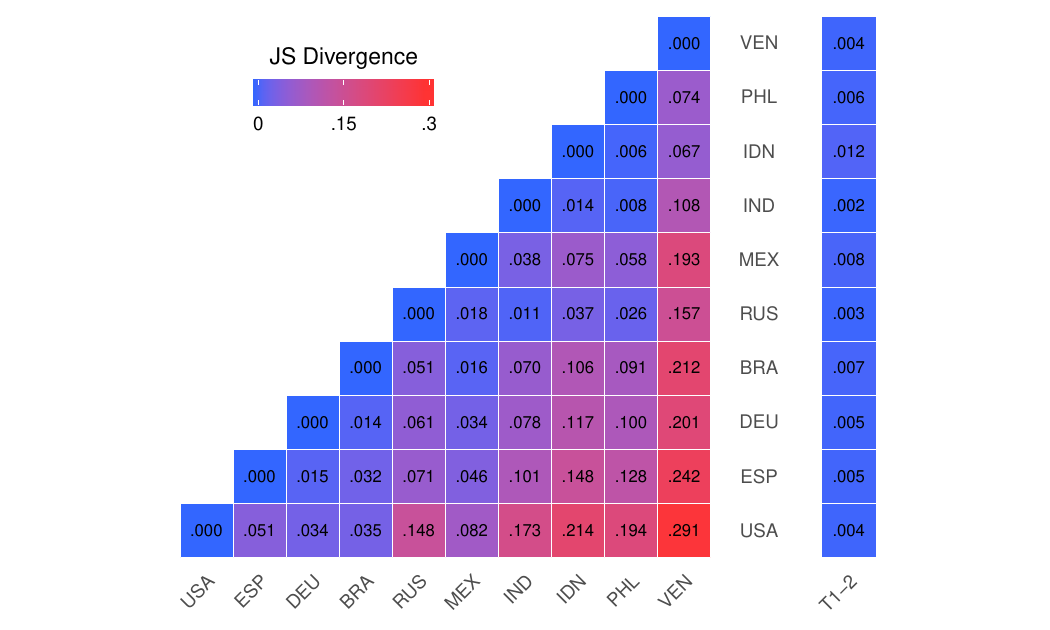}
  \caption{\textbf{Yearly Household Income JS}. This figure shows the JS divergences between the household income distributions of the different countries. The bar on the right shows the JS divergence between \texttt{T1} and \texttt{T2} for each individual country.}
  \label{fig:household_income_js}
\end{figure}

Regarding the difference between the two time points, we found the largest differences in Russia, Brazil, and Venezuela. Russia had fewer workers with a Bachelor's or Master's degree in \texttt{T2} than in \texttt{T1}. In Brazil, the proportion of workers reporting a high school degree but no college degree increased, and the proportion of workers reporting some high school, a Bachelor's degree, or associate degree decreased. 
In Venezuela, the proportion of high school graduates with no college increased, while the proportion of workers reporting vocational training or an associate degree decreased.

\subsection{Yearly Household Income}

In order to meaningfully capture household income in a set of countries with wildly varying average incomes, we created logarithmic bins\footnote{We rounded the logarithmically spaced numbers for better readability in the answer options.} for the response options. The question asked workers to report an estimate of their annual disposable household income (i.e., after taxes) in US dollars. Figure~\ref{fig:household_income} shows the household income distribution for each country.

While Venezuela was classified as an ``upper middle income'' country by the World Bank during the time of data collection \cite{WorldBankClassification}, workers from Venezuela reported by far the lowest annual household income.
Apart from Venezuela, the reported income distributions are largely consistent with the World Bank classification of the countries, with the United States, Spain, and Germany reporting higher incomes (despite the smaller reported household size) and India, Indonesia, and the Philippines reporting lower incomes.
Unsurprisingly, data from mturk tracker\footnote{\url{http://demographics.mturk-tracker.com/\#/householdIncome/all}} shows that on \amt, Indian workers also tend to report lower household incomes than workers from the United States.

While the proportion of workers reporting an annual income below US\$ 3,000 was much higher in low income countries than in high income countries, a significant proportion of workers in high income countries also reported a yearly household income of less than US\$ 3,000. There might be several explanations for this, such as students living on student loans, unemployed workers living off their savings, or workers on welfare benefits who do not consider the benefits as ``income.''

Figure~\ref{fig:household_income_js} shows the differences between the household income distributions.
The largest differences in household income were generally found between the countries in the low income group (and Venezuela) and the countries in the high income group, with the largest difference being between the USA and Venezuela. 

Between \texttt{T1} and \texttt{T2}, the household income distributions remained largely stable. We observed the largest change in Indonesia, where in \texttt{T2} more workers reported a yearly income below US\$ 3,000 (39\%) than in \texttt{T1} (30\%).
The second-largest change between the time points was in Mexico, where the number of workers reporting a household income between US\$32,000 and US\$50,000 decreased from 12\% to 7\% while the proportion of workers reporting a lower income increased.

\section{Importance of Microtasks for Crowdworkers}
\label{sec:importance}

In this section, we compare the importance of microtasks and microtask income for workers in the ten different countries. Our survey included three questions about different aspects concerning the centrality of microtasks in the workers' lives (see questions I1-I3 in Table~\ref{table:survey_questions}). Analogously to the previous section, we report the proportion of each answer choice in the ten countries as an average between \texttt{T1} and \texttt{T2} as well as the JS divergences \cite{lin1991divergence} of the answer distributions between the countries and between the two time points. The results presented in this section are also available as an interactive visualization on our website \cite{crowdworkersinfo}.

\subsection{Weekly Time Spent on CrowdFlower}

\begin{figure}[t]
  \centering
    \includegraphics[width=0.9\textwidth]{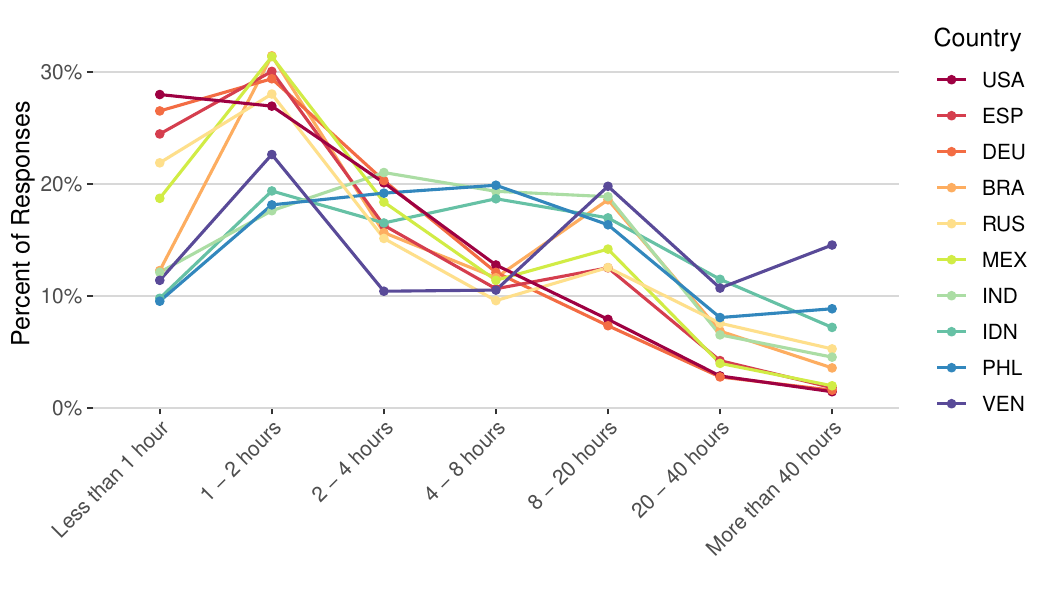}    \vspace*{-7mm}
  \caption{\textbf{Time Spent on CrowdFlower per Week.} This figure shows the distribution of weekly time spent on the platform by workers in the different countries. The percentages represent the average of \texttt{T1} and \texttt{T2}. Venezuela, the Philippines, and Indonesia were the countries with the highest proportion of workers who reported spending more than 20 hours per week on CrowdFlower. Venezuela had the highest proportion of workers spending more than 40 hours per week on the platform.}
  \label{fig:weekly_time}
\end{figure}

Figure~\ref{fig:weekly_time} shows, for the ten countries, how much time workers report spending on CrowdFlower per week.
Venezuela, the Philippines, and Indonesia were the countries with the highest proportion of workers who reported spending more than 20 hours per week on CrowdFlower, and Venezuela had the highest proportion of workers spending more than 40 hours per week on the platform.
In all countries, but especially in the countries in the high and middle income groups, there was a significant proportion of workers who used CrowdFlower less than two hours per week. The countries in the high income group had the highest proportion of workers who reported spending less than one hour per week on CrowdFlower.

\begin{figure}[h!]
  \centering
\includegraphics[width=0.8\textwidth]{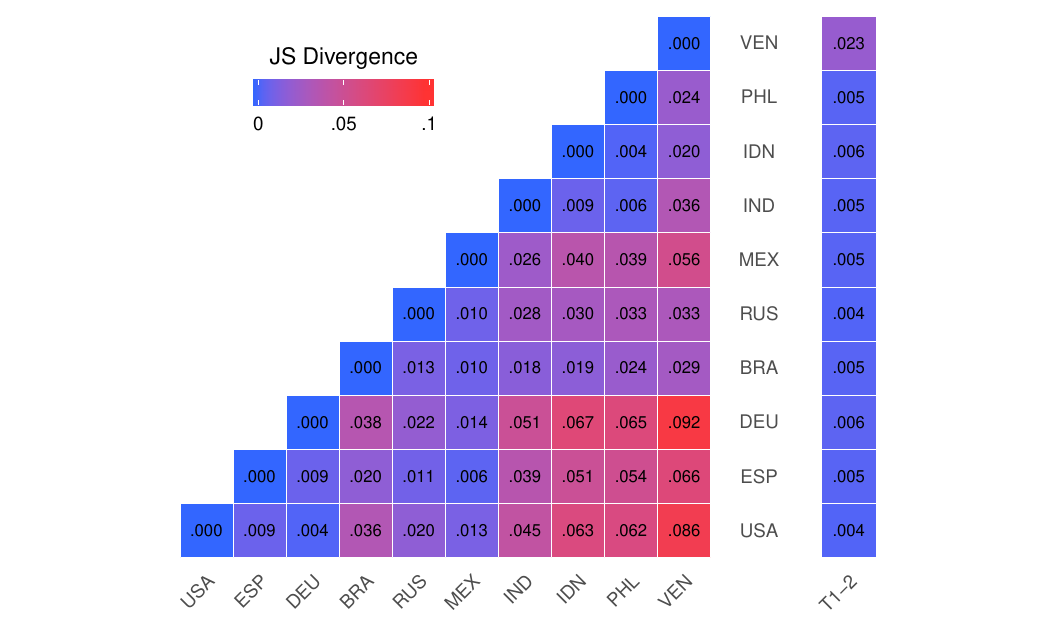}
  \caption{\textbf{JS of Weekly Time Spent on CrowdFlower}. This figure shows the JS divergences between the answer distributions of the different countries. The bar on the right shows the JS divergence between \texttt{T1} and \texttt{T2} for each individual country.}
  \label{fig:weekly_time_js}
\end{figure}

Figure~\ref{fig:weekly_time_js} shows the JS divergences between the answer distributions.
Regarding the differences between countries, countries in the high income group were generally most dissimilar to countries in the low income group, with countries in the low income group generally spending more time on the platform.

The largest change in distribution between \texttt{T1} and \texttt{T2} was in Venezuela. In \texttt{T2}, the proportion of Venezuelan workers spending over 40 hours per week on CrowdFlower (19.5\%) was almost double the proportion reported in \texttt{T1} (9.6\%). This increase was likely due to changes in the country's economic situation, making CrowdFlower an increasingly attractive source of income for Venezuelan workers.

\subsection{Dependency on Microtask Income}

\begin{figure}
  \centering
    \includegraphics[width=0.5\textwidth]{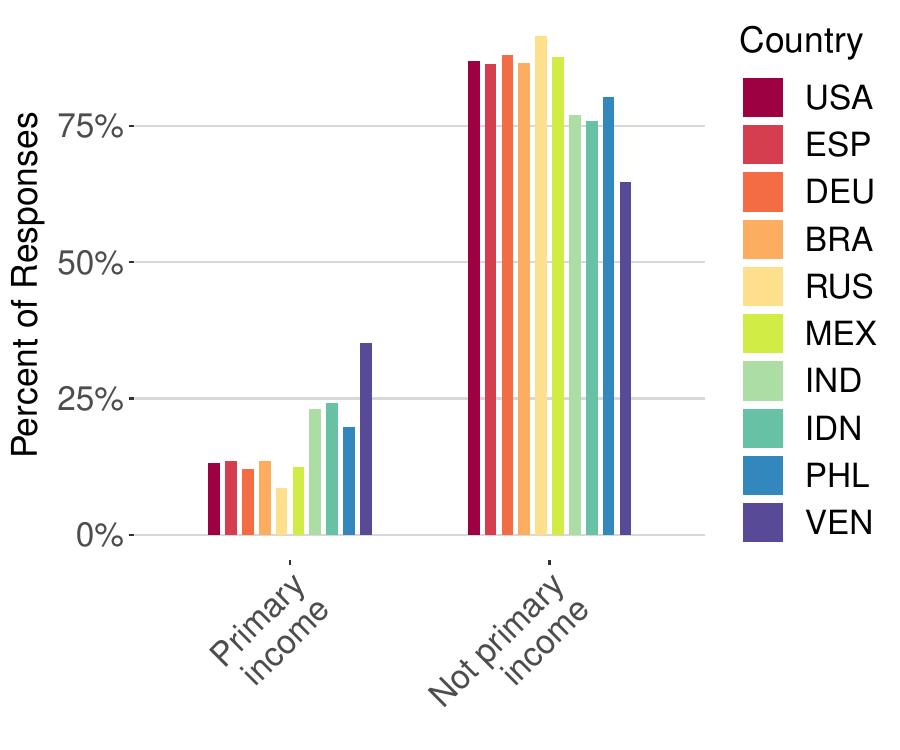}
    \caption{\textbf{Dependency on CrowdFlower Income.} This figure shows the proportion of workers who reported microtask income being their primary/non-primary source of income in the different countries. The bar heights represent the averages of \texttt{T1} and \texttt{T2}. In countries of the low income group as well as in Venezuela, a higher percentage of crowdworkers reported relying on CrowdFlower as their primary source of income.}
    \label{fig:primary_income}
\end{figure}

Crowdworkers in countries of the high and middle income groups reported the lowest percentages of reliance on CrowdFlower as their primary source of income. There were no large differences between the countries of the high income group and those of the middle income group, and the lowest reliance on CrowdFlower as a main source of income was in Russia, a country in the middle income group. 
In the low income group as well as in Venezuela, the proportions were significantly higher. 
Figure~\ref{fig:primary_income} shows the answer distribution of each country.

\begin{figure}
  \centering
    \includegraphics[width=0.8\textwidth]{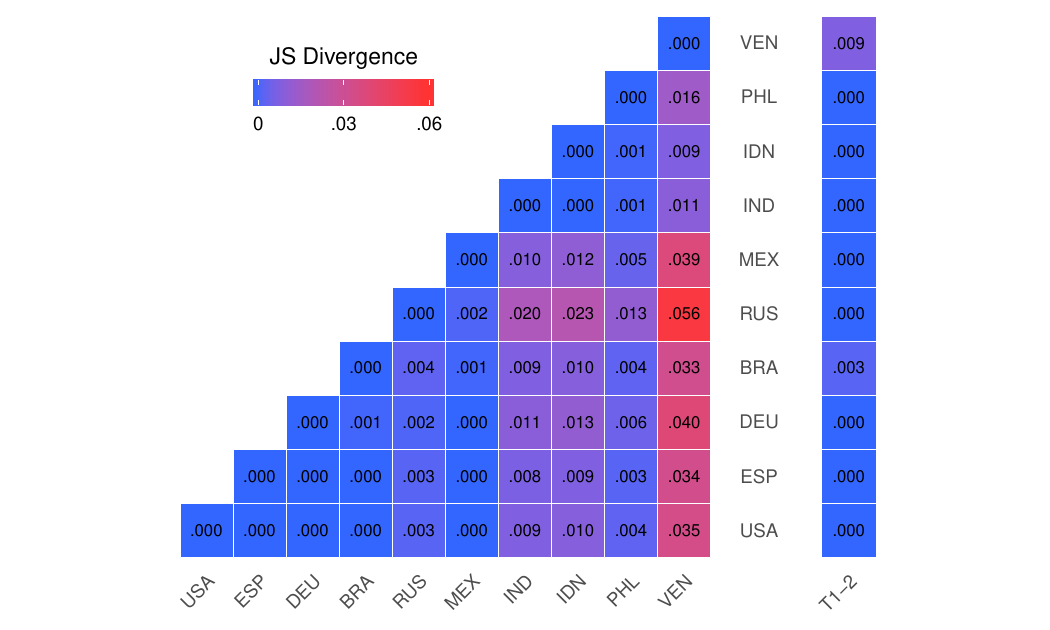}
    \caption{\textbf{Dependency on CrowdFlower Income JS.} This figure shows the JS divergences between the answer distributions of the different countries. The bar on the right shows the JS divergence between \texttt{T1} and \texttt{T2} for each individual country.}
    \label{fig:primary_income_js}
\end{figure}

In terms of distribution differences between countries, Venezuela had the highest JS divergences with other countries, especially with the countries in the high and middle income group. The countries in the high and middle income categories were very similar to each other. Figure~\ref{fig:primary_income_js} shows the JS divergences of the answer distributions.

Our results regarding American and Indian workers are similar to early findings on \amt.
For example, \citet{ipeirotis2010demographics} found that about 15\% of American workers and less than 30\% of Indian workers reported \amt as their primary source of income. 
However, later studies on \amt reported higher percentages: \citet{berg2016income} found that 38\% of American workers and 49\% of Indian workers on \amt relied on crowdwork as their primary source of income, and \citet{brawley2016work} reported 39\% and 41\% for American and Indian workers, respectively. This suggests that, at the time of our data collection, CrowdFlower was less likely than \amt to be used as a main source of income by workers from these countries. A reason for this might be that workers can achieve a higher hourly income on \amt than on CrowdFlower (see, e.g., \citealp{berg2016income}).

\FloatBarrier

The reliance of workers on CrowdFlower as a main source of income was mostly stable between \texttt{T1} and \texttt{T2}, with the exception of Venezuela and, to a lesser extent, Brazil. In Venezuela, consistent with the increase of weekly time spent on the platform, the percentage of workers relying on CrowdFlower as a primary source of income significantly increased from \texttt{T1} (29\%) to \texttt{T2} (41.5\%). In Brazil, the percentage was also higher in \texttt{T2} (15.9\%) than in \texttt{T1} (11.1\%).

\subsection{Use of Microtask Income}

\begin{figure}[t!]
  \centering
     \includegraphics[width=0.99\textwidth]{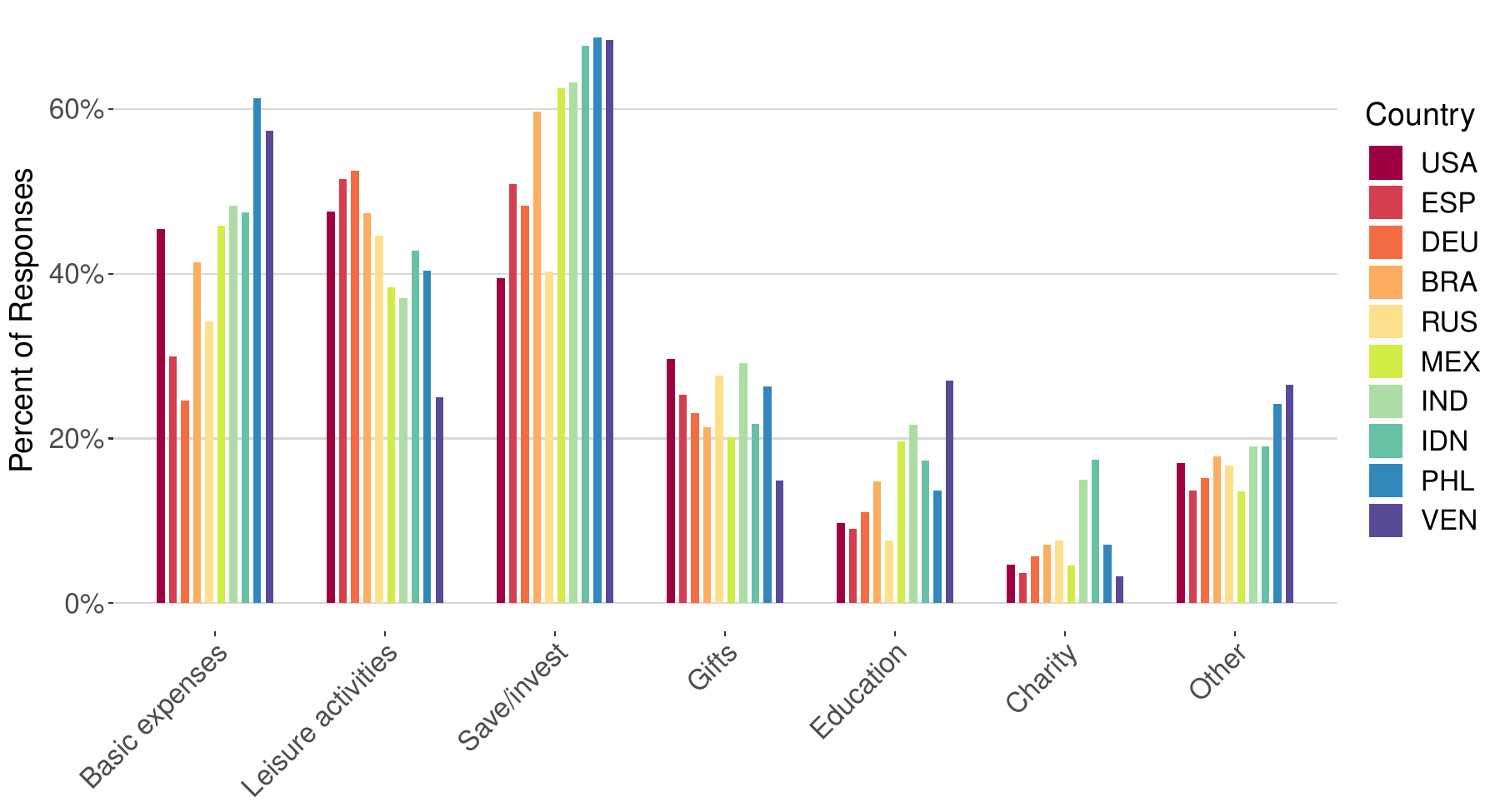}
  \caption{\textbf{Use of CrowdFlower Income.} This figure shows how workers spend their income from microtasks in the different countries. The bar heights represent the averages of \texttt{T1} and \texttt{T2}. In seven out of ten countries, the proportion of workers who reported spending their microtask income on basic expenses such as food, rent, sanitary items, or medical care exceeded 40\%.}
  \label{fig:money_use}
\end{figure}

The question regarding workers' use of the income earned through microtasks offered seven answer options (see Table~\ref{table:freetext_money_use}), and workers could select one or more of the options.
Figure~\ref{fig:money_use} shows the proportion of workers who selected the different expenditure categories, for each country.\footnote{Note that the sum of the different categories may be higher than 100\% for each country, as workers could choose more than one expenditure category.}

In seven out of ten countries, the proportion of workers who reported spending microtask income on basic expenses such as food, rent, sanitary items, or medical care exceeded 40\%. The countries with the highest proportion of workers who spent the money on basic expenses were the Philippines and Venezuela.
Germany was the country with the lowest percentage of workers spending the money on basic expenses, followed by Spain and Russia.
In the USA, despite being a high income country, over 40\% of workers reported spending the money on basic expenses. 

The three countries in the high income group and Brazil had the highest percentage of workers who stated spending the money on leisure activities such as hobbies, games, holidays, or sports. In all other countries except Venezuela, the proportion of workers who reported spending microtask income on leisure activities was also higher than 30\%.
In Venezuela, the proportion of workers who reported spending microtask income on leisure activities was by far lowest of all countries.

A high percentage of crowdworkers indicated that they save or invest the money earned on CrowdFlower, especially in lower income countries. The countries where the highest percentage of workers chose this response were Venezuela, the Philippines, and Indonesia. The USA and Russia had the lowest proportions of workers who reported saving or investing the income from microtasks. 

The USA, Russia, and India had the highest proportion of workers who reported spending the money on gifts, while the lowest proportion for this expenditure category was in Venezuela.
A moderate percentage of workers stated using the microtask income for financing their education. This expenditure category was highest in Venezuela, followed by India, Mexico, and Indonesia. 
In most countries, very few workers donate their income from microtasks to charities, with the exception of India and Indonesia.  
A significant proportion of workers also stated that they used the money for purposes other than the given categories, especially in the Philippines and Venezuela.

\begin{figure}
  \centering
     \includegraphics[width=.8\textwidth]{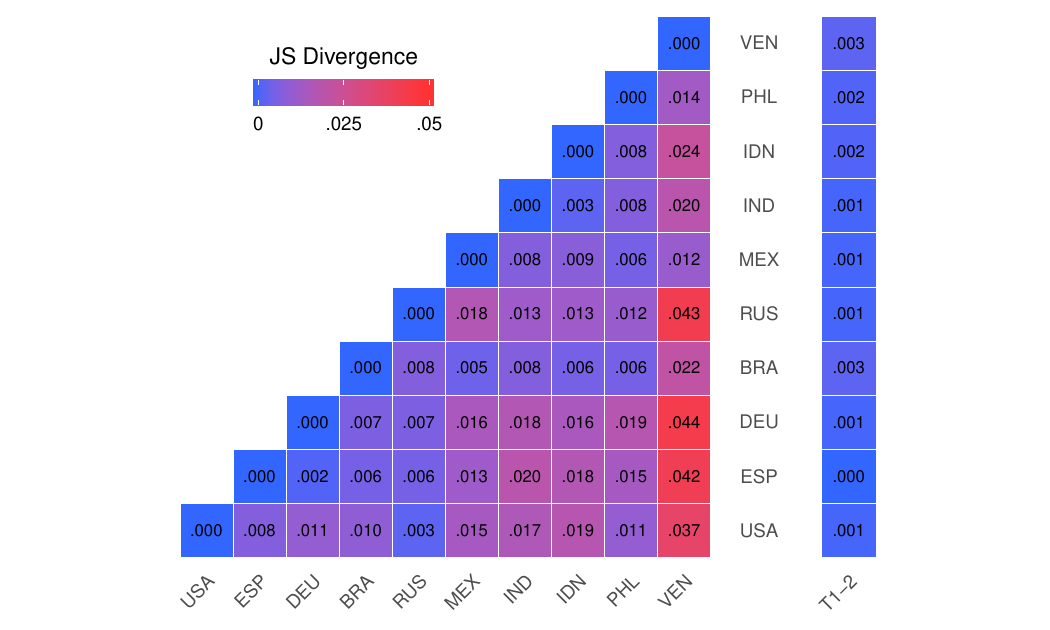}
  \caption{\textbf{Use of CrowdFlower income JS.} This figure shows the JS divergences between the answer distributions of the different countries. The bar on the right shows the JS divergence between \texttt{T1} and \texttt{T2} for each individual country.}
  \label{fig:money_use_js}
\end{figure}

\FloatBarrier

Figure~\ref{fig:money_use_js} shows the JS divergences of the answer distributions. As this survey question allowed for multiple answers, we normalized the distributions to sum to one before calculating the JS divergences.
We found the largest differences in distribution between Venezuela and the three countries in the high income group as well as Russia. Generally, the countries in the high income group, as well as Russia, were somewhat similar among each other, and more dissimilar to the countries in the low income group and Venezuela.

Regarding the difference between the time points, Venezuela showed the largest changes. These changes were mostly in the categories \emph{basic expenses} and \emph{education}.
While in \texttt{T1} the country with the highest proportion of workers spending the microtask money for basic expenses was the Philippines, in \texttt{T2} it was Venezuela. The proportion of Venezuelan workers who reported using the microtask income for basic expenses rose from 52.4\% in \texttt{T1} to 62.4\% in \texttt{T2}. This is consistent with Venezuelan workers' increase in relying on CrowdFlower as a primary source of income and with their increase in weekly time spent on the platform.
The proportion of Venezuelan workers using the money for their education rose from 22\% at \texttt{T1} to 32\% at \texttt{T2}.

The second-largest change\footnote{Brazil had a slightly lower JS (0.0025) than Venezuela (0.0028).} from \texttt{T1} to \texttt{T2} was observed in Brazil. In \texttt{T2}, fewer Brazilian workers indicated saving or investing their microtask income, while more Brazilian workers reported spending it on basic expenses and leisure activities. 
In \texttt{T2}, there was also a lower percentage of workers who reported donating microtask income to charity than in \texttt{T1}, in all countries of the low income group.

\section{Discussion}
\label{sec:discussion}

To gain a better understanding of the international crowd workforce, this paper
has presented a comprehensive analysis of the demographic characteristics of workers in ten countries and of the centrality of microtask income in workers' lives.
This section summarizes and discusses the principal findings of our study.

\subsection{Differences Between Countries} 
The results of our study revealed many substantial differences between the crowd workforces of different countries, both regarding the different demographic characteristics and regarding the importance of income from microtasks.

Regarding the demographic composition of the crowd workforce, we found substantial differences between countries in all characteristics, but the patterns of differences did not necessarily correspond to the countries' income groups for all characteristics.
For example, our results showed that workers were predominantly female in the USA and predominantly male in most other countries. Only workers from the Philippines had a nearly balanced gender distribution. Crowdworkers were generally young, but workers in Russia tended to be older than workers from other countries, and they were also more likely to be married than workers in other countries. Regarding the workers' employment status, a substantial proportion of workers in all countries reported having a full-time job besides their activity on CrowdFlower. With the exception of Venezuela, over 35\% of workers in all countries reported having a full-time job.

The demographic characteristics for which the pattern of country differences seemed to correspond to country income groups were household income, household size, and education level. The differences in household income and household size are not surprising: Crowdworkers in low income countries reported lower household incomes and larger household sizes than crowdworkers in high income countries. More interesting are the differences in education level: While crowdworkers in all countries tended to be well-educated, countries in the low income group generally had a higher proportion of workers with academic degrees than high income countries. For example, workers from India reported a much higher percentage of Master's degrees than high income countries, while workers in the Philippines had a very high percentage of workers with a Bachelor's degree. Indonesia also had a high proportion of workers with a Bachelor's degree. Venezuela, classified as an upper-middle income country by the World Bank \cite{WorldBankClassification} but with workers reporting a much lower household income than even workers in countries of the low income group, had an education level distribution that was closer to countries of the high (and middle) income group.

Overall, these results imply that researchers using microtasks in their projects, as well as task requesters from industry, should be aware of the fact that the crowd workforces in different countries do not only have varying cultural backgrounds, but that they may also have vastly differing distributions of other characteristics such as gender or education level. When a microtask is offered to workers without any country restrictions, the results may therefore contain different kinds of biases depending on the country distribution of the workers. It is especially important to keep this in mind when comparing the results of tasks that were worked on by varying proportions of workers from different countries. The results obtained from our analysis of demographic characteristics can inform researchers and other task requesters about which demographic groups they are likely to be targeting by offering a task to workers in a specific country. We recommend that social and behavioral science researchers who use CrowdFlower (or similar microtask platforms) to sample research participants always gather comprehensive data on participants' demographic characteristics. This enables researchers to detect -- and if necessary statistically account for -- cross-national differences in the demographic composition of the crowdworker samples.

Our results also showed substantial differences between countries regarding the role that microtask income plays in the workers' lives. 
Workers in the high income group reported spending less time on CrowdFlower than workers in the low income group. Venezuela had the largest proportion of workers who reported spending more than 40 hours per week on the platform. In high income countries, a much larger percentage of workers reported spending less than two hours per week on the platform than in low income countries. These results are congruent with our findings regarding the dependency on microtask income. A lower percentage of workers in high and middle income countries reported relying on microtask income than in low income countries, and there were no large differences between the percentages of the middle and the high income group. Venezuela was the country where most crowdworkers relied on microtask income as their primary source of income. 
\enlargethispage{-1\baselineskip}

However, it is important to note that microtask income not being a worker's primary income does not mean that the income has little significance for the worker. In fact, in seven out of the ten countries, over 40\% of workers reported spending the income from CrowdFlower on basic expenses such as food, rent, sanitary items, or medical care. The only countries where this percentage was lower than 40\% (but still higher than 20\%) were Spain, Germany, and Russia. The countries with the highest proportion of workers who reported spending money on basic expenses were the Philippines and Venezuela. Notably, over 40\% of workers in the USA, despite it being a high income country, reported spending the money on basic expenses.

Overall, the results regarding the importance of microtask income show that the income from microtasks is of high significance for many workers and that a large proportion of workers may rely on this income to help pay for basic expenses such as food or rent. Platform designers and task requesters should be aware of this fact when making decisions that affect the workers' experience. For example, when designing platform functionality
such as processes for suspending worker accounts, platform designers
should make sure that these processes are transparent and that
there is a possibility for workers to appeal unfair decisions. For task
requesters, it is important to provide clear task instructions and to ensure that their quality control mechanisms are fair. It is also important that a task's implementation is well-tested before offering the task to workers. Failing to keep these aspects in mind may result in workers not being paid for completed work, in many cases threatening their
livelihoods. Finally, the results of our survey also have implications for policy makers, as they provide information on which demographic groups any policies and regulations regarding work on microtask platforms are likely to concern, as well as how new policies and regulations might impact the crowdworkers' lives.

\subsection{Stability of the Characteristics}

To gain insights into the stability of the different characteristics, we repeated our survey after eight months and compared the results to the first round of data collection. The results of this comparison showed that the workforce's characteristics were mostly stable between the two large samples taken independently at different points in time.

The countries that showed substantial changes in several characteristics were Venezuela and, to a lesser extent, Brazil.
In Venezuela, the characteristics that showed notable changes between \texttt{T1} and \texttt{T2} were the weekly time spent on the platform, the dependency on the income from the platform, and the use of the income. Regarding the weekly time spent on the platform, the percentage of Venezuelan workers who reported spending over 40 hours on CrowdFlower almost doubled from \texttt{T1} (9.6\%) to \texttt{T2} (19.5\%). The proportion of workers who depended on income from CrowdFlower as their primary source of income also greatly increased from \texttt{T1} (29\%) to \texttt{T2} (41.5\%), and the percentage of Venezuelan workers who reported spending the income on basic expenses rose from 52.4\% in \texttt{T1} to 62.4\% in \texttt{T2}. There was also a change in the proportion of Venezuelan workers who reported having a full-time job besides their activity on Crowdflower, with the percentage dropping from 28\% in \texttt{T1} to 23\% in \texttt{T2}.

These changes in the Venezuelan crowd workforce reflect the exceptional economic situation that Venezuela was in at the time of our data collection. According to \citet{IMF.gdp} estimates, Venezuela's GDP contracted by 17\% in 2016 and by 15.7\% in 2017. The country also suffered from extremely high inflation throughout the period of data collection. In November 2016, Venezuela entered a state of hyperinflation, with the inflation rate exceeding 50\% per month \cite{hanke2016venezuela}. 

Brazil had the largest change in the employment status distribution of all countries. Specifically, there was a large decrease of workers who reported being in full-time employment (from 59.5\% in \texttt{T1} to 46.2\% in \texttt{T2}) and a large increase of unemployed workers (from 13.1\% in \texttt{T1} to 21.6\% in \texttt{T2}). 
Brazil also had the second-largest change in the percentage of workers who relied on the income from CrowdFlower as their main source of income, with the percentage rising from 11.1\% in \texttt{T1} to 15.9\% in \texttt{T2}. Furthermore, the country had the second-largest change in the distribution of money use: In \texttt{T2}, a lower percentage of workers reported that they saved or invested their microtask income, while a higher percentage reported spending it on basic expenses and leisure activities.
Like the change in Venezuela, the change in Brazil might be the result of changing economic circumstances in the country during our period of data collection. Except for Venezuela, Brazil was the only country in our study where the GDP per capita contracted in 2016 \cite{worldbank.gdp}. Also, unemployment in Brazil had been sharply rising since 2015, and it continued to rise throughout 2016 and 2017 \cite{worldbank.unemployment}.

Apart from these exceptions, however, the characteristics were largely stable between the two samples. This stability also indicates that our samples were large enough to accurately capture the workforce's characteristics. Of course, this does not mean that the characteristics of crowd workforces in different countries will be stable indefinitely. For example, in the American workforce on the platform \amt, a gradual decrease in female workers has been observed over time (see, e.g., \citealp{ross2010crowdworkers}). However, in our study, most characteristics were stable in most countries, and the largest changes that we observed could be explained by a changing economic situation in the respective countries.

\section{Conclusion}
\label{sec:conclusion}

The work presented in this paper constitutes the first comparison of crowdworker characteristics at the country level that both i) goes beyond an analysis of the two countries that constitute the majority of workers on \amt and ii) has been conducted at a large scale. By shedding light on the country-specific differences of the international crowd workforce, this study complements existing research and contributes to a better understanding of this emerging form of work.

We presented an analysis of the demographic composition of the crowd workforce in ten countries and the centrality of microtask income in workers' lives. We based our analysis on two large samples of crowdworkers from ten different countries, collected at two different points in time on the platform CrowdFlower.
Our results reveal significant differences in demographic composition, time spent on the platform, reliance on microtask income as well as use of microtask income between the different countries. Furthermore, our results show that the characteristics of the workforce in different countries remained, in most cases, largely stable between the two samples collected eight months apart. While there were changes in the answer distributions of certain characteristics in some countries, the average differences between the countries were larger than the average change over time. These results constitute an important step towards a more comprehensive characterization of the international crowd workforce.
They can also help -- apart from insights into the microtask labor market in and by itself -- to inform researchers posting similar tasks on the studied platform as to what kind of audience they might potentially reach with certain selection settings, especially such related to nationality. 

Our study has several limitations. While we took great care to account for fluctuations in worker composition (e.g., by the hour of the day or the day of the week) by dividing the starting times of our tasks into different categories, further research on the daily and weekly fluctuations of the different characteristics is needed. Furthermore, due to the nature of microtasks, our samples are necessarily self-selected. Our samples therefore do not include workers who, for example, exclusively work on repeatable tasks and never accept survey tasks. Lastly, our sample focuses on workers who have sufficient English skills to understand the survey questions. However, this is likely true for the majority of the microtask workforce on this platform, as workers are expected to understand instructions in English\footnote{The platform's interface is available exclusively in English.} and demand for crowdworkers is driven by Anglophone countries \cite{kuek2015global}.

In future work, we plan to analyze the relationship between demographic characteristics and motivational profiles of crowdworkers, using the Multidimensional Crowdworker Motivation Scale \cite{mcms_published}. 
Future research will also be able to use the data presented in this study in order to compare the demographic composition of the crowd workforce with the composition of the general population, and the general workforce, in different countries.
Furthermore, future work is encouraged to study the characteristics of workers on platforms other than CrowdFlower and \amt, of which
there is still little knowledge, especially at the country level. For collecting data on the characteristics of workers on other platforms, the survey developed in this paper can be easily adapted.
To further deepen our understanding of the international crowd workforce, future research is also encouraged to investigate why the demographic composition of the crowd workforce differs between countries, and what factors lead to a country's workforce being disproportionately composed of a certain demographic.
Finally, future research focusing on the examination of factors that cause differences in crowd workforce composition over time will further contribute to a better understanding of the phenomenon of crowdwork. 
This paper is relevant for researchers, practitioners, and policy makers interested in the composition of the international crowd workforce. Further information about our work is available on our website \cite{crowdworkersinfo}.

%%% -*-BibTeX-*-
%%% Do NOT edit. File created by BibTeX with style
%%% ACM-Reference-Format-Journals [18-Jan-2012].

\DeclareRobustCommand{\firstsecond}[2]{#2}

\appendix
\pagebreak

\section{Survey Questions}
\label{app:answer_options}

\newcommand{\tab}{\hspace{5mm}} % used in Item Table

\centering
\begin{longtable}{p{8mm} p{15mm} l} 
\caption{\textbf{Survey Questions and Answer Options.} This table shows all survey questions and answer options used in our CrowdFlower task.} \\
\toprule
\textbf{Demographics} & & \\
\toprule
  \mbox{D1: \it{What is your gender?}} & \\
  & D1.O1 & Male\\ 
  & D1.O2 & Female\\
  & D1.O3 & Other\\%[1mm]   
\midrule
  \mbox{D2: \it{What is your age?}} & \\
  & D2.O1 & Under 18 years old\\ 
  & D2.O2 & 18 to 24 years\\ 
  & D2.O3 & 25 to 34 years\\
  & D2.O4 & 35 to 44 years\\    
  & D2.O5 & 45 to 54 years\\
  & D2.O6 & 55 to 64 years\\   
  & D2.O7 & Age 65 or older\\    
\midrule
  \mbox{D3: \it{What is your marital status?}} & \\
  & D3.O1 & Single (never married)\\ 
  & D3.O2 & Married\\ 
  & D3.O3 & Separated\\
  & D3.O4 & Widowed\\    
  & D3.O5 & Divorced\\
\midrule
  \mbox{D4: \it{How many people live in your household?}} & \\
  & D4.O1 & 1 person\\ 
  & D4.O2 & 2 persons\\ 
  & D4.O3 & 3 persons\\
  & D4.O4 & 4 persons\\ 
  & D4.O5 & 5 persons\\
  & D4.O6 & 6 persons\\ 
  & D4.O7 & More than 6 persons\\     
\midrule
  \mbox{D5: \it{What is your highest education level?}} & \\
  & D5.O1 & No schooling completed\\ 
  & D5.O2 & Some high school\\ 
  & D5.O3 & High school graduate\\
  & D5.O4 & Some college, no degree\\    
  & D5.O5 & Trade/technical/vocational training\\
  & D5.O6 & Associate degree\\   
  & D5.O7 & Bachelor's degree\\   
  & D5.O8 & Master's degree\\   
  & D5.O9 & Doctorate degree\\ 
 \pagebreak
\midrule
  \mbox{D6: \it{What is your employment status (CrowdFlower tasks excluded)?}} & \\
  & D6.O1 & Unemployed\\ 
  & D6.O2 & In education\\ 
  & D6.O3 & Part-time\\
  & D6.O4 & Full-time\\    
  & D6.O5 & Retired\\
  & D6.O6 & Other\\   
\midrule
  \mbox{D7: \makecell[tl]{\it{What is your approximate household income, per YEAR (after taxes,}\\  \it{in US\$)?}}} & \\
  & D7.O1 & Less than \$3,000\\ 
  & D7.O2 & \$3,000 - \$5,000\\ 
  & D7.O3 & \$5,000 - \$8,000\\
  & D7.O4 & \$8,000 - \$12,000\\    
  & D7.O5 & \$12,000 - \$20,000\\
  & D7.O6 & \$20,000 - \$32,000\\   
  & D7.O7 & \$32,000 - \$50,000\\   
  & D7.O8 & \$50,000 - \$80,000\\   
  & D7.O9 & \$80,000 - \$125,000\\   
  & D7.O10 & More than \$125,000\\     
\toprule
\textbf{\mbox{Importance of Microtasks}} & \\
\toprule
  \mbox{I1: \it{How much time do you spend on CrowdFlower, per week?}} & \\
  & I1.O1     & Less than 1 hour\\ 
  & I1.O2     & 1 - 2 hours\\   
  & I1.O3     & 2 - 4 hours\\  
  & I1.O4     & 4 - 8 hours\\ 
  & I1.O5     & 8 - 20 hours\\
  & I1.O6     & 20 - 40 hours\\    
  & I1.O7     & More than 40 hours\\ 
\midrule
  \mbox{I2: \it{Is the money from CrowdFlower your primary source of income?}} & \\
  & I2.O1 & Yes\\ 
  & I2.O2 & No\\
\midrule
  \mbox{I3: \it{What do you do with the money that you earn on CrowdFlower?}} & \\
  & I3.O1     & \makecell[tl]{I use the money for basic living expenses (food, rent, \\ \tab sanitary items, medical care, ...).}\\ 
  & I3.O2     & \makecell[tl]{I spend the money on leisure activities (hobbies, games, \\ \tab  holidays, sports, ...).}\\   
  & I3.O3     & I save/invest the money.\\  
  & I3.O4     & I use the money to buy gifts for other people.\\ 
  & I3.O5     & I use the money to finance my education.\\
  & I3.O6     & I donate the money to charity.\\    
  & I3.O7     & Other purposes.\\  
\bottomrule
\label{table:mcms_items}
\end{longtable}

\end{document}